\documentclass[aps,prd,reprint,longbibliography,footinbib,bibnotes,superscriptaddress]{revtex4-2}
\usepackage{amsmath,amssymb,mathrsfs}
\usepackage{xcolor}
\usepackage{graphicx,tikz}
\usepackage{CJKutf8}
\usepackage[colorlinks=true,allcolors=blue!55!black]{hyperref}
\newcommand{\F}{\mathcal F}
\newcommand{\Hh}{\mathcal H}
\newcommand{\N}{\mathcal N}
\newcommand{\Z}{\mathbf Z}
\newcommand{\M}{\mathbf M}
\newcommand{\T}{\mathbf T}
\newcommand{\PT}{\operatorname{PT}}
\newcommand{\tr}{\operatorname{tr}}
\newcommand{\Sp}{\operatorname{Sp}}
\newcommand{\LP}{\operatorname{LP}}
\newcommand{\dd}{\mathrm d}
\newcommand{\ii}{\mathrm i}
\newcommand{\sym}{\mathcal S}
\newcommand{\nc}{N_c}
\newcommand{\eps}{\epsilon}
\newcommand{\Tcal}{\mathcal T}
\newcommand{\Jcal}{\mathcal J}
\begin{document}

\title{Bootstrapping the Full-Color Two-Loop Four-Point NMHV Form Factor}
\hypersetup{
  pdftitle={Bootstrapping the Full-Color Two-Loop Four-Point NMHV Form Factor},
  pdfauthor={Jiahao Liu}
}

\author{Jiahao Liu (刘家昊)}
\email{liujiahao@itp.ac.cn}
\affiliation{New Cornerstone Laboratory, Institute of Theoretical Physics,\\
Chinese Academy of Sciences, Beijing 100190, China}
\affiliation{School of Physical Sciences, University of Chinese Academy of Sciences,\\
No.~19A Yuquan Road, Beijing 100049, China}

\begin{abstract}
We compute the symbol and complete function of the full-color two-loop
four-point NMHV form factor of the chiral stress-tensor multiplet in
$\N=4$ super-Yang--Mills theory after minimal infrared subtraction.
The complete hard function uses 148 letters, each matching one of the 78
letters in the planar two-loop NMHV result up to permutation.
In overlapping three-particle channels, double discontinuities vanish
and the symbol satisfies extended Steinmann relations.
From the triple-collinear and double-soft limits, we extract the
corresponding two-loop hard splitting and soft contributions to the NMHV
form factor, including subleading color.
\end{abstract}

\begin{CJK*}{UTF8}{gbsn}
\maketitle
\end{CJK*}

\section{Introduction}
\label{sec:intro}

Form factors are matrix elements of local operators between the vacuum
and on-shell states. The momentum entering through the operator can be
off shell, introducing an additional scale into the scattering problem.
In $\N=4$ super-Yang--Mills theory, the chiral stress-tensor multiplet
has been a setting for developing superspace and on-shell methods, from
early perturbative calculations to generalized unitarity and on-shell
diagrams \cite{BorkKazakovVartanov:2011,Brandhuber:2011,Bork:2013,Frassek:2016}.
In the planar theory, bootstrap methods have reached eight loops for the
three-point form factor \cite{DixonEtAl:2022EightLoops}, and the periodic
Wilson-loop description has extended two-loop calculations of maximally
helicity-violating (MHV) form factors to higher multiplicities
\cite{Li:2025PeriodicWilsonLoop}.
Studies of planar form factors have revealed antipodal self-duality of
the four-point MHV remainder \cite{DixonSelfDual:2023} and shown how
cluster algebras organize form-factor symbol alphabets
\cite{ChicherinHennPapathanasiou:2021,HeLiuCluster:2026}.

The planar two-loop four-point MHV form factor is known at function
level \cite{GuoEtAl:2025MHV,DixonXin:2025}. The full-color result has
been computed by combining numerical unitarity with a bootstrap
\cite{HeJiang:2026} and, independently, by integrating a representation
with global color--kinematics duality \cite{CaiGuoYang:2026}.
These studies also give subleading-color results for triple-collinear
splitting and double-soft emission. In the next-to-maximally
helicity-violating (NMHV) sector, the planar four-point form factor
is known through three loops at the symbol level \cite{HeLiuYang:2026}.
Here we determine the symbol and the complete function of the full-color
two-loop four-point NMHV hard form factor in a minimal infrared
subtraction scheme.

We first construct a symbol ansatz with kinematic
prefactors from four-dimensional leading singularities and a scalar
basis from one-mass pentagon functions \cite{Chicherin:2022,Abreu:2024}.
After imposing symmetry and physical constraints, we require a single
set of two-loop integral coefficients for which the infrared poles cancel
and the finite symbol matches the ansatz.
We independently compute the full-color one-loop four-point NMHV form factor
through $O(\eps^2)$ and use it in the infrared subtraction.
The remaining symbol coefficients are fixed by $D$-dimensional
unitarity cuts. We then use the full master-integral expansions and
single-soft factorization to determine the terms beyond the symbol.
After matching the subtraction schemes, our leading-color symbol
agrees with the planar result \cite{HeLiuYang:2026}. Each full-color letter
matches a planar two-loop letter under relabeling. The complete hard
function has vanishing double discontinuities in overlapping
three-particle channels, and its symbol satisfies extended Steinmann
relations in those channels. We use the triple-collinear and double-soft
limits to determine the corresponding two-loop hard splitting and soft
contributions to the NMHV form factor.

\section{Kinematics and Color Decomposition}
\label{sec:conventions}
We consider the chiral part of the stress-tensor supermultiplet,
$\mathcal O(x,\theta^+)$, in $\N=4$ super-Yang--Mills theory with
gauge group $\mathrm{SU}(N_c)$. Its lowest component is the half-BPS
operator $\operatorname{tr}(\phi^2)$. The four-point super form factor of
$\mathcal O$ is
\cite{Brandhuber:2011,HeLiuYang:2026}
\begin{equation}
\begin{split}
\F_4(1,2,3,4;q,\gamma)
 &=\int\dd^4x\,\dd^4\theta^+\,
     e^{-\ii(q\cdot x+\theta^+\cdot\gamma)}\\
 &\quad\times\langle 1,2,3,4|\mathcal O(x,\theta^+)|0\rangle .
\end{split}
\label{eq:form-factor}
\end{equation}
Each label $i=1,\ldots,4$ denotes an on-shell superstate with outgoing
massless momentum $p_i$ and Grassmann coordinates $\eta_i$.
We write $q=\sum_i p_i$ and $s_I=(\sum_{i\in I}p_i)^2$.
Below, $\F_4$ is stripped of the overall delta functions
$\delta^{(4)}(\sum_i p_i-q)\delta^{(8)}(Q-\gamma)$, where
$Q=\sum_i\lambda_i\eta_i$ and $\gamma$ is the chiral operator
supermomentum. The stripped NMHV coefficient has Grassmann degree four.

We work in dimensional reduction ($D=4-2\eps$) and set $\mu^2=q^2=1$.
The hard form factor is obtained by minimal infrared subtraction,
\begin{equation}
\Hh_4=\lim_{\eps\to0}\Z^{-1}_4\F_4 ,
\label{eq:hard-def}
\end{equation}
where $\Z_4$ is the pure-pole infrared operator in color space, determined
through two loops by the dipole anomalous dimension
\cite{Becher:2009,Gardi:2009}.
Appendix~\ref{app:subtraction} gives $\Z_4$ and defines the coupling $a$.
At two loops, Eq.~\eqref{eq:hard-def} gives
\begin{equation}
\medmuskip=2mu\relax
\begin{split}
\Hh^{(2)}_4
 =\lim_{\eps\to0}\big[\F^{(2)}_4-\Z^{(1)}_4\F^{(1)}_4
+\big((\Z^{(1)}_4)^2-\Z^{(2)}_4\big)\F^{(0)}_4\big].
\end{split}
\label{eq:hard}
\end{equation}

We use fundamental generators normalized by
$\tr(T^aT^b)=\delta^{ab}$ and write
$\tr(i_1\cdots i_k)=\tr(T^{a_{i_1}}\cdots T^{a_{i_k}})$,
where $a_i$ is the adjoint index of leg $i$.
For $\sigma_i=(abcd)$ with $\sigma_1=(1234)$, $\sigma_2=(1243)$,
and $\sigma_3=(1324)$, the trace tensors are
\[
S_i=\tr(abcd)+\tr(adcb),\quad D_i=2\tr(ac)\tr(bd).
\]
The four-adjoint color relations give the two-loop NMHV decomposition
\cite{Naculich:2012,NaculichOsathapan:2024,HeJiang:2026}
\begin{equation}
\begin{split}
\Hh^{(2)}_{4,\mathrm{NMHV}}
 &=\sum_{i=1}^3\big[S_i(\nc^2\mathcal A_i+\mathcal C_i)
          +\nc D_i\mathcal B_i\big],\\
\sum_i\mathcal C_i&=0,\qquad
\mathcal B_i=\sum_j\mathcal A_j-\tfrac12\mathcal C_i .
\end{split}
\label{eq:trace}
\end{equation}

\section{The Bootstrap}
\label{sec:bootstrap}

\subsection{Leading singularities and function space}

We obtain the kinematic prefactors from four-dimensional on-shell
diagrams built from tree-level amplitudes and form factors.
The rational prefactors are products of the ordered Parke--Taylor factor
$\PT(ijkl)=1/(\langle ij\rangle\langle jk\rangle
\langle kl\rangle\langle li\rangle)$ and NMHV form-factor $R$-invariants
in periodic momentum-twistor variables \cite{Bork:2013,Bork:2014,KosterEtAl:2017}.
We write $i^\pm=i\pm4$ for the periodic images.
The two representatives for the $(1234)$ ordering are
\begin{align}
R_1={}&[2,3,4,4^-,1],\nonumber\\
\bar R_1={}&
\frac{\langle1\,3^-\,4^-\,4\rangle\langle1\,3\,4\,3^-\rangle}
{\langle1^+\,3\,4\,1\rangle\langle4\,4^-\,3\,3^-\rangle}
[3^-,4^-,3,4,1].
\label{eq:Rreps}
\end{align}
Here $[a,b,c,d,e]$ is the NMHV five-bracket and
$\langle abcd\rangle=\det(Z_a,Z_b,Z_c,Z_d)$ is the momentum twistor
four-bracket. Cyclic relabeling gives the remaining
$R_i$ and $\bar R_i$. For each ordering $\sigma$, let $V_\sigma$
be the span of $\PT(\sigma)R_i^{(\sigma)}$ and
$\PT(\sigma)\bar R_i^{(\sigma)}$ for $i=1,\ldots,4$.
The invariants satisfy the alternating relation
\begin{equation}
\sum_{i=1}^4(-1)^{i-1}(R_i+\bar R_i)=0 .
\label{eq:Rrel}
\end{equation}
Each $V_{\sigma_i}$ has dimension 7 over $\mathbb Q$, and their sum
$V_{\sigma_1}+V_{\sigma_2}+V_{\sigma_3}$, denoted by $V_{\rm rat}$,
has dimension 13. For the rational prefactors of $\mathcal A_i$, we use
the 7 Parke--Taylor-dressed invariants in ordering $\sigma_i$ obtained
by omitting $\bar R_4$. For the independent subleading single-trace
coefficients $\mathcal C_1$ and $\mathcal C_2$, we extend the $(1234)$
basis by the Parke--Taylor-dressed $R_1,\ldots,R_4,\bar R_1,\bar R_3$
in the $(1243)$ ordering.

The algebraic prefactors come from three-mass triangle cuts
\cite{Bork:2013,BianchiEtAl:2019}.
For the partition $ab|cd$, define
\begin{equation}
\begin{split}
\Delta_{ab|cd}&=\sqrt{(q^2-s_{ab}-s_{cd})^2-4s_{ab}s_{cd}},\\
z_\pm&=\frac{q^2+s_{ab}-s_{cd}\pm\Delta_{ab|cd}}{2q^2}.
\end{split}
\end{equation}
The values $z_\pm$ solve $(zq-p_a-p_b)^2=0$. We write the corresponding
null momenta as $z_\pm q-p_a-p_b=\lambda_\pm\tilde\lambda_\pm$ and define
\begin{align}
U(z_\pm)&=\frac{\delta^{(4)}\bigl(z_\pm\langle\lambda_\pm Q\rangle
                   -\langle\lambda_\pm a\rangle\eta_a
                   -\langle\lambda_\pm b\rangle\eta_b\bigr)}
                {z_\pm(1-z_\pm)\langle ab\rangle\langle cd\rangle
                     \prod_{j=1}^4\langle\lambda_j,\lambda_\pm\rangle},\\
\Tcal_{ab|cd}&=\frac{U(z_+)+U(z_-)}{2\Delta_{ab|cd}}.
\label{eq:triangle}
\end{align}
Changing the sign of $\Delta_{ab|cd}$ exchanges $z_+$ and $z_-$, so
$\Tcal_{ab|cd}$ changes sign. The triangle prefactors in $\mathcal A_1$,
$\mathcal A_2$, and $\mathcal A_3$ are $(\Tcal_{12|34},\Tcal_{14|23})$,
$(\Tcal_{12|34},\Tcal_{13|24})$, and $(\Tcal_{13|24},\Tcal_{14|23})$,
respectively. The two prefactors in each pair multiply separate functions
in the ansatz. In the independent subleading coefficients
$(\mathcal C_1,\mathcal C_2)$, each triangle prefactor multiplies a common
function with relative weights $(1,1)$, $(-2,1)$, and $(1,-2)$ for
$12|34$, $13|24$, and $14|23$, respectively. The coefficient $\mathcal C_3$
and the double-trace contributions follow from Eq.~\eqref{eq:trace}. This gives
$3\cdot7+2\cdot13=47$ rational and $6+3=9$ algebraic color-dressed
prefactors. The $\nc^2$ and $\nc^0$ single traces account for 27 and 29
of them, respectively.

We take the scalar functions from the full two-loop
one-mass pentagon-function space
\cite{Chicherin:2022,Abreu:2024,Abreu:2023data,AbreuItaMoriello:2020}.
To use these data we identify the five momenta with
$(-q,p_1,p_2,p_3,p_4)$ and take the reference production channel
with $p_3,p_4$ incoming. The external Gram root is oriented by
$\operatorname{tr}_5=4\ii\varepsilon(p_1,p_2,p_3,p_4)$.
At fixed $q^2$, the weight-four symbol ansatz uses 6468
pentagon-function monomials $F_a$.
We label the 56 color-dressed kinematic prefactors by $\mathcal K_b$
and denote the symbol map on scalar functions by $\sym$. The ansatz is
\begin{equation}
\sym(\Hh^{(2)}_{4,\mathrm{NMHV}})
 =\sum_{b=1}^{56}\mathcal K_b\sum_{a=1}^{6468}c_{ba}\,\sym(F_a),
 \quad c_{ba}\in\mathbb Q .
\label{eq:ansatz}
\end{equation}

\subsection{Constraints on the symbol ansatz}

We impose crossing symmetry by requiring the color-dressed ansatz to be
invariant under relabeling of the four external legs.
The Gram root $\operatorname{tr}_5$ changes sign under odd permutations. For the
Galois constraints, we hold $\operatorname{tr}_5$ fixed and require
invariance under a sign flip of any one of the other square roots
in the pentagon alphabet. The symbol multiplying each $\mathcal K_b$
must acquire the same sign as the prefactor. The Steinmann relations
\cite{Steinmann:1960a,Steinmann:1960b,CaronHuot:2016} require the double
discontinuity to vanish for every ordered pair
$i\ne j$ of overlapping three-particle channels $v_i=(q-p_i)^2/q^2$.
At symbol level, we replace the first two letters of each word by the
product of their $\dd\log$ residues in Eq.~\eqref{eq:steinmann-residues}
and require the resulting sum to vanish. At the representative
spurious pole $\langle3|q|4]=0$, we require the sum of prefactor
residues weighted by the corresponding boundary symbols to vanish.
We impose this cancellation separately for each power of the logarithm
in the expansion near the pole. Crossing extends this constraint to
all $\langle i|q|j]=0$ with $i\ne j$.

At leading power, the double-collinear and single-soft limits of the
color-dressed hard form factor give
\begin{equation}
\medmuskip=1mu\relax
\begin{split}
\underset{i\parallel j}{\LP}\,\Hh_4^{(2)}
 &=\Sp_{ij,H}^{(2)}\Hh_3^{(0)}
   +\Sp_{ij,H}^{(1)}\Hh_3^{(1)}
   +\Sp_{ij,H}^{(0)}\Hh_3^{(2)},\\[2pt]
\underset{p_i\to0}{\LP}\,\Hh_4^{(2)}
 &=\Jcal_{i,H}^{(2)}\Hh_3^{(0)}
   +\Jcal_{i,H}^{(1)}\Hh_3^{(1)}
   +\Jcal_{i,H}^{(0)}\Hh_3^{(2)}.
\end{split}
\label{eq:single-limits}
\end{equation}
For the splitting amplitude $X=\Sp_{ij}$ or the soft current
$X=\Jcal_i$, we use $X_H=\lim_{\eps\to0}\Z_4^{-1}X\Z_3$, where $\Z_3$ is
the corresponding infrared subtraction operator for the three-leg form factor.
We construct the right-hand sides from the known three-point form factor
\cite{Brandhuber:2012}, splitting amplitudes \cite{BernDixonKosower:2004},
and full-color soft current \cite{DixonSoft:2020}, using the subtraction
scheme of Eq.~\eqref{eq:hard-def}. After stripping the common delta
functions, we match the symbols of the terms of Grassmann degree four
to the leading-power limits of the ansatz in each color component.

We expand the bare two-loop form factor in pure scalar integrals,
leaving their coefficients free in each color component. We add the
infrared subtraction in Eq.~\eqref{eq:hard}, using the one-loop form factor
obtained from the integrand in Appendix~\ref{app:one-loop}.
The double pole of $\Z_4^{(1)}$ requires the one-loop form factor
through $O(\eps^2)$. We retain the symbol of the maximal-weight part
at each order in $\eps$ and require a single set of integral
coefficients for which the poles cancel and the finite symbol matches
the ansatz. Eliminating the integral coefficients gives linear
constraints on the hard symbol.

Table~\ref{tab:counts} lists the number of free parameters in the
symbol ansatz after each step of the bootstrap.

\subsection{\texorpdfstring{$D$}{D}-dimensional unitarity cuts}

We evaluate the eight-, seven-, and six-propagator cuts shown in
Appendix~\ref{app:cut-topologies} over finite fields
\cite{Bern:1994unitarity,Giele:2008,Peraro:2016,Abreu:2017unitarity}.
We extract the $\mathrm{SU}(4)$-invariant Grassmann coefficients from the
twelve components with two identical scalars and two gluons of opposite
helicity.
We check these coefficients at two independent kinematic points using
the prescription of Appendix~\ref{app:one-loop}. We reduce the cut integrals
to uniform-weight master integrals with \texttt{Kira} 3.1
\cite{Maierhofer:2018,Lange:2026}.

The cut equations fix the remaining symbol coefficients.
As an independent check of the reconstructed symbol, we use
a four-propagator cut evaluated at fixed transverse directions, together
with the corresponding one-loop subtraction. Our
leading-color symbol in the $(1234)$ ordering agrees with the planar
NMHV result of Ref.~\cite{HeLiuYang:2026} after matching
conventions~\footnote{%
For the comparison we strip the planar powers of $N_c$ at each loop
order and express the prefactors in a common convention. The
conversion to the planar subtraction scheme adds
$K\Hh^{(1)}+\tfrac12 K^2\Hh^{(0)}$ to our two-loop hard function at
symbol level, where $K=\tfrac12\sum_{i=1}^4\log^2(s_{i,i+1}/q^2)$
with cyclic indices.}.

\begin{table}[t]
\caption{\label{tab:counts}Remaining symbol parameters.}
\centering
\begin{minipage}{0.84\columnwidth}
\begin{ruledtabular}
\begin{tabular}{lr}
Step & Parameters\\
\hline
Initial ansatz & $362\,208$\\
Crossing symmetry & $15\,593$\\
Galois symmetry & $10\,943$\\
Steinmann relations & $7\,214$\\
Spurious-pole cancellation & $1\,664$\\
Double-collinear limit & $598$\\
Single-soft limit & $446$\\
Integral and infrared matching & $104$\\
Numerical Unitarity & $0$
\end{tabular}
\end{ruledtabular}
\end{minipage}
\end{table}

\subsection{The complete hard function}
\label{sec:uplift}

With the symbol fixed, we replace $\sym(F_a)$ in
Eq.~\eqref{eq:ansatz} by the corresponding pentagon-function
monomials on the production sheet \cite{Abreu:2024,Abreu:2023data}.
To constrain terms beyond the symbol, such as zeta values
multiplying lower-weight functions, we match the full $\eps$
expansions of the pure master integrals to the infrared poles
and the finite symbol, including the one-loop subtraction in
Eq.~\eqref{eq:hard}.

After imposing crossing and Galois symmetry, we fix the remaining terms
beyond the symbol using single-soft factorization in
Eq.~\eqref{eq:single-limits}. The resulting hard function
satisfies the eight-, seven-, and six-propagator cut equations through
the finite term and agrees with double-collinear factorization.

\section{The Result}
\label{sec:analytic}

We determine the letter support from the iterated-integral
expansion of the complete hard function, including the terms
beyond the symbol. The full-color result uses 148 letters, each related
by an external-leg permutation to one of the 78 letters in the planar
two-loop NMHV result \cite{HeLiuYang:2026}.

For each ordered pair of overlapping three-particle channels,
we continue from the production sheet to a neighborhood of
$v_i=v_j=0$. For $k=i,j$, we write $\operatorname{Disc}_{v_k}$ for the
change in the function after one counterclockwise circuit around $v_k=0$.
We find
\begin{equation}
\operatorname{Disc}_{v_j}\operatorname{Disc}_{v_i}\,
\Hh^{(2)}_{4,\mathrm{NMHV}}=0,\qquad i\ne j .
\label{eq:doublemon}
\end{equation}
This verifies the Steinmann relations
in all color components, including the terms beyond the symbol.

To test the extended Steinmann relations \cite{CaronHuot:2019},
we replace each adjacent pair $W_k\otimes W_{k+1}$ in the symbol by
\begin{equation}
\bigl(\operatorname{Res}_{v_i=0}\dd\log W_k\bigr)
\bigl(\operatorname{Res}_{v_j=0}\dd\log W_{k+1}\bigr),
\label{eq:steinmann-residues}
\end{equation}
with residues taken on the continued sheet. The resulting sum
vanishes for every $i\ne j$ and $k=1,2,3$. The cases $k=2,3$
are checks on the reconstructed symbol.

\section{Triple-collinear and double-soft limits}
\label{sec:limits}

We take the triple-collinear and double-soft limits of the full-color hard
form factor using the pentagon functions for the decay channel
\cite{ChenEtAl:2026FourJet,ChenEtAl:2026data}. Both limits factorize onto
the two-point Sudakov form factor.
Removing the terms that contain its loop corrections isolates the
two-loop hard splitting amplitude or double-soft current acting on the
tree-level two-point form factor.

Write $V_\ell=\LP[\rho\,\Hh_4^{(\ell)}]$, where $1/\rho$ is the
leading power in either limit.
For $X=\Sp$ or $\Jcal$, define $X_H=\lim_{\eps\to0}\Z_4^{-1}X\Z_2$, where
$\Z_2$ is the infrared subtraction operator for the Sudakov form factor.
With $\Hh_2^{(\ell)}=\nc^\ell h_\ell\Hh_2^{(0)}$, the extracted
two-loop term is
\begin{equation}
 \rho X_H^{(2)}\Hh_2^{(0)}
 =V_2-\nc h_1 V_1+\nc^2(h_1^2-h_2)V_0 .
\label{eq:boundary}
\end{equation}
The coefficients $h_1$ and $h_2$ are obtained from the Sudakov form
factor \cite{Gehrmann:2012Sudakov} and are given in Eq.~\eqref{eq:sudakov}.

\subsection{Triple-collinear splitting}

We take $p_2,p_3,p_4$ collinear to a parent of momentum $P$ and set
$\rho=s_{234}/q^2$. As $\rho\to0$, we hold fixed
\begin{equation}
 z_i=\frac{s_{1i}}{q^2-s_{234}},\qquad
 a_{ij}=\frac{s_{ij}}{s_{234}},\qquad i,j\in\{2,3,4\},
\label{eq:colshape}
\end{equation}
with $\sum_i z_i=\sum_{i<j}a_{ij}=1$ and fixed Gram orientation. The
$z_i$ become the longitudinal momentum fractions. Taking $X=\Sp$ in
Eq.~\eqref{eq:boundary} gives the two-loop $1\to3$ hard splitting
amplitude contracted with $\Hh_2^{(0)}$ in the NMHV sector.

The tree-level two-point form factor identifies the color indices of $P$
and leg $1$, so we use the trace decomposition in Eq.~\eqref{eq:trace}.
The rational prefactors have collinear limits
\begin{equation}
\begin{split}
K_{\sigma,R_i}&=\LP_{\rho\to0}
  \bigl[\rho\,\PT(\sigma)R_i^{(\sigma)}\bigr],\\
K_\sigma^{\rm tree}&=K_{\sigma,R_1}+K_{\sigma,R_3}-K_{\sigma,\bar R_2} .
\end{split}
\label{eq:collinear-prefactors}
\end{equation}
The same definition applies to $K_{\sigma,\bar R_i}$. The second line gives
the tree-level splitting prefactor $K_\sigma^{\rm tree}$.
The collinear limits of the prefactors in $V_\sigma$ and $V_{\rm rat}$
span spaces of dimensions 5 and 8 over $\mathbb Q$, respectively.
For each ordering we use the basis
$\mathscr K_\sigma=\{K_\sigma^{\rm tree},K_{\sigma,R_1},K_{\sigma,R_2},
K_{\sigma,R_3},K_{\sigma,\bar R_1}\}$.
For $\sigma=(1,i,j,k)$, the two triangle prefactors in the corresponding
leading single trace reduce to
\begin{equation}
\medmuskip=1mu\relax
\begin{split}
2\LP\bigl[\rho\,\Tcal_{1i|jk}\bigr]
 &=K_\sigma^{\rm tree}-K_{\sigma,R_2}
   +K_{\sigma,R_3}+K_{\sigma,\bar R_1},\\
2\LP\bigl[\rho\,\Tcal_{1k|ij}\bigr]
 &=K_{\sigma,\bar R_1}-K_{\sigma,R_1} .
\end{split}
\label{eq:collinear-triangles}
\end{equation}
The triangle roots are taken with $\Delta_{1a|bc}\to q^2(1-z_a)$.
The subleading single traces share the basis
$\mathscr K_{\rm rat}=\mathscr K_{(1234)}\cup
\{K_{(1243)}^{\rm tree},K_{(1243),R_1},K_{(1243),R_2}\}$.
The same identities express their triangle prefactors in this basis.

For the leading NMHV term of $\rho\,\Sp_H^{(2)}\Hh_2^{(0)}$,
the single-trace coefficients in Eq.~\eqref{eq:trace} are
\begin{equation}
\begin{split}
\mathcal A_i^{\rm Sp}&=\sum_{K\in\mathscr K_{\sigma_i}} K f_{i,K},
  \qquad i=1,2,3,\\
\mathcal C_i^{\rm Sp}&=\sum_{K\in\mathscr K_{\rm rat}} K\tilde f_{i,K},
  \qquad i=1,2.
\end{split}
\label{eq:collinear-splitting}
\end{equation}
The third subleading single trace and all double traces follow from
the relations in Eq.~\eqref{eq:trace}.
The weight-four functions $f_{i,K}$ and $\tilde f_{i,K}$ depend on the ratios
in Eq.~\eqref{eq:colshape} and are polynomials in $\log\rho$ of degree at most 4.
The four coefficients of
$K_{(1234)}^{\rm tree}$ and $K_{(1234),R_1}$ in $\mathcal A_1^{\rm Sp}$
and $\mathcal C_1^{\rm Sp}$ determine the others by permutations of
the collinear legs and parity.

\subsection{Double-soft current}

We take $p_1,p_2$ soft with $p_3,p_4$ hard and set
$\rho=s_{12}/q^2$. The limiting ratios are
\begin{equation}
\begin{aligned}
&\frac{(s_{13},s_{14},s_{23},s_{24})}{q^2\sqrt{\rho}}
 \longrightarrow \frac{(A,B,C,D)}{\sqrt b},\\
&b=AD+BC-2\sqrt{ABCD}\cos\varphi .
\end{aligned}
\label{eq:softshape}
\end{equation}
Here $A,B,C,D$ are positive and sum to 4.
The angle $\varphi$ is the oriented azimuthal separation of the soft
momenta in the plane transverse to $p_3,p_4$. The sign of $\sin\varphi$
specifies the Gram orientation. The expression for $b$ follows from
$s_{34}/q^2\to1$ and the identity
\begin{equation*}
s_{12}s_{34}=s_{13}s_{24}+s_{14}s_{23}
 -2\sqrt{s_{13}s_{14}s_{23}s_{24}}\cos\varphi ,
\end{equation*}
obtained by decomposing the soft momenta along and transverse to $p_3,p_4$.

We define the soft prefactors by
\begin{equation}
\begin{split}
K^{\rm soft}_{\sigma,R}&=\LP_{\rho\to0}
  \bigl[\rho\,\PT(\sigma)R^{(\sigma)}\bigr],\\
\Tcal^{\rm soft}_{ab|cd}&=\LP_{\rho\to0}\bigl[\rho\,\Tcal_{ab|cd}\bigr].
\end{split}
\label{eq:soft-prefactors}
\end{equation}
For the $(1234)$ and $(1243)$ orderings, the seven prefactors from
$R_1,\ldots,R_4,\bar R_1,\bar R_2,\bar R_3$ give bases of the same rational
space. For $(1324)$ the basis consists of $K^{\rm soft}_{(1324),R_2}$
and $K^{\rm soft}_{(1324),R_4}$. We denote the basis for each ordering by
$\mathscr S_\sigma$ and use $\mathscr S_{\rm rat}=\mathscr S_{(1234)}$ for
both independent subleading single traces.
The two mixed soft--hard triangle prefactors have the same leading limit,
which lies in the rational space. For $\sigma=(1234)$ or $(1243)$, the relation is
\begin{equation}
\begin{split}
2\Tcal^{\rm soft}_{13|24}=2\Tcal^{\rm soft}_{14|23}
={}&-K^{\rm soft}_{\sigma,R_2}-K^{\rm soft}_{\sigma,R_3}\\
 &+K^{\rm soft}_{\sigma,\bar R_2}+K^{\rm soft}_{\sigma,\bar R_3} .
\end{split}
\label{eq:soft-triangles}
\end{equation}
For $\sigma=(1324)$, the right-hand side is
$K_{\sigma,R_2}^{\rm soft}+K_{\sigma,R_4}^{\rm soft}$.
Only $\Tcal^{\rm soft}_{12|34}$ remains independent of the rational prefactors, with
$\Delta_{12|34}/(q^2\sqrt\rho)\to\sqrt{16/b-4}$.

Taking $X=\Jcal$ in Eq.~\eqref{eq:boundary} gives
$\rho\,\Jcal_H^{(2)}\Hh_2^{(0)}$. Its NMHV single-trace coefficients in
Eq.~\eqref{eq:trace} are
\begin{equation}
\begin{split}
\mathcal A_i^{\rm soft}&=\sum_{K\in\mathscr S_{\sigma_i}} K g_{i,K}
  +\Tcal_{12|34}^{\rm soft}g_{i,\triangle},\quad i=1,2,3,\\
\mathcal C_i^{\rm soft}&=\sum_{K\in\mathscr S_{\rm rat}} K\tilde g_{i,K}
  +\Tcal_{12|34}^{\rm soft}\tilde g_{\triangle},\quad i=1,2 .
\end{split}
\label{eq:soft-current}
\end{equation}
The color factors of $\Tcal_{12|34}$ give $g_{3,\triangle}=0$ and a common
coefficient $\tilde g_{\triangle}$ for $\mathcal C_1^{\rm soft}$ and
$\mathcal C_2^{\rm soft}$. The third subleading single trace and the double
traces follow from Eq.~\eqref{eq:trace}. The weight-four functions depend
on $A,B,C,\cos\varphi$ at fixed Gram orientation and are polynomials in
$\log\rho$ of degree at most 4. The $\nc^0$ single-trace terms have degree
at most 2.
The remaining coefficients follow from the nine representatives supplied
in the ancillary files by exchanging the two soft legs, exchanging the
two hard legs, and applying parity.

\section{Conclusions and outlook}

We have obtained the symbol and complete function of the full-color
two-loop four-point NMHV hard form factor of the chiral stress-tensor
multiplet in $\N=4$ super-Yang--Mills theory. Each of the 148 letters in
the complete function matches one of the 78 letters in the planar
two-loop NMHV result up to permutation.
In overlapping three-particle channels, double discontinuities of the
complete function vanish in every color component, and the symbol
satisfies the extended Steinmann relations. From the triple-collinear and
double-soft limits, we extract the NMHV contributions of the two-loop
hard splitting amplitude and double-soft current acting on the tree-level
two-point form factor.

The MHV integrand of Ref.~\cite{CaiGuoYang:2026} satisfies global
color--kinematics duality and motivates a similar construction for the
NMHV form factor. It would also be interesting to test whether the letter
support and Steinmann relations persist at higher loops.
At three points, the symbol of the two-loop form factor remainder agrees
with the weight-four symbol of the leading-color $H\to ggg$ remainder
after matching the infrared subtraction and normalization
\cite{Brandhuber:2012,GehrmannEtAl:2012Higgs3Partons}. This motivates a
comparison at four points \cite{HeJiang:2026}.
The two-loop leading-color Higgs-plus-two-jet
amplitudes in the heavy-top effective theory
\cite{DeLaurentisEtAl:2026Higgs2Jets} provide a starting point. One can
compare the corresponding gluon helicity components at weight four,
with the operator component and infrared subtraction matched, and
test whether any agreement extends to subleading color.

\begin{CJK*}{UTF8}{gbsn}
\section*{Acknowledgements}
The author would like to thank Song He \mbox{(何颂)} and Xuhang Jiang
\mbox{(姜旭航)} for helpful discussions. OpenAI Codex assisted with literature
searches, exploratory calculations, and language editing. The author is
responsible for the scientific content of this work. This work was supported
by the New Cornerstone Science Foundation.
\end{CJK*}

\section*{Ancillary Files}
The ancillary files provide the symbol and complete pentagon-function
expression for the full-color two-loop four-point NMHV hard form factor.
For the triple-collinear and double-soft limits, we give the symbols and
iterated-integral expressions for the coefficient functions defined in
Sec.~\ref{sec:limits}. The accompanying README explains how to load the
Mathematica files and reconstruct the results from the supplied
representative functions.

\makeatletter
\renewcommand{\bibsection}{%
  \par
  \begingroup
    \baselineskip26pt
    \bib@device{\hsize}{72pt}%
  \endgroup
  \nobreak\@nobreaktrue
  \addvspace{19pt}%
}
\makeatother
\bibliography{references}

\begin{thebibliography}{44}%
\makeatletter
\providecommand \@ifxundefined [1]{%
 \@ifx{#1\undefined}
}%
\providecommand \@ifnum [1]{%
 \ifnum #1\expandafter \@firstoftwo
 \else \expandafter \@secondoftwo
 \fi
}%
\providecommand \@ifx [1]{%
 \ifx #1\expandafter \@firstoftwo
 \else \expandafter \@secondoftwo
 \fi
}%
\providecommand \natexlab [1]{#1}%
\providecommand \enquote  [1]{``#1''}%
\providecommand \bibnamefont  [1]{#1}%
\providecommand \bibfnamefont [1]{#1}%
\providecommand \citenamefont [1]{#1}%
\providecommand \href@noop [0]{\@secondoftwo}%
\providecommand \href [0]{\begingroup \@sanitize@url \@href}%
\providecommand \@href[1]{\@@startlink{#1}\@@href}%
\providecommand \@@href[1]{\endgroup#1\@@endlink}%
\providecommand \@sanitize@url [0]{\catcode `\\12\catcode `\$12\catcode
  `\&12\catcode `\#12\catcode `\^12\catcode `\_12\catcode `\%12\relax}%
\providecommand \@@startlink[1]{}%
\providecommand \@@endlink[0]{}%
\providecommand \url  [0]{\begingroup\@sanitize@url \@url }%
\providecommand \@url [1]{\endgroup\@href {#1}{\urlprefix }}%
\providecommand \urlprefix  [0]{URL }%
\providecommand \Eprint [0]{\href }%
\providecommand \doibase [0]{https://doi.org/}%
\providecommand \selectlanguage [0]{\@gobble}%
\providecommand \bibinfo  [0]{\@secondoftwo}%
\providecommand \bibfield  [0]{\@secondoftwo}%
\providecommand \translation [1]{[#1]}%
\providecommand \BibitemOpen [0]{}%
\providecommand \bibitemStop [0]{}%
\providecommand \bibitemNoStop [0]{.\EOS\space}%
\providecommand \EOS [0]{\spacefactor3000\relax}%
\providecommand \BibitemShut  [1]{\csname bibitem#1\endcsname}%
\let\auto@bib@innerbib\@empty
\bibitem [{\citenamefont {Bork}\ \emph {et~al.}(2011)\citenamefont {Bork},
  \citenamefont {Kazakov},\ and\ \citenamefont
  {Vartanov}}]{BorkKazakovVartanov:2011}%
  \BibitemOpen
  \bibfield  {author} {\bibinfo {author} {\bibfnamefont {L.~V.}\ \bibnamefont
  {Bork}}, \bibinfo {author} {\bibfnamefont {D.~I.}\ \bibnamefont {Kazakov}},\
  and\ \bibinfo {author} {\bibfnamefont {G.~S.}\ \bibnamefont {Vartanov}},\
  }\bibfield  {title} {\bibinfo {title} {{On form factors in N=4 sym}},\ }\href
  {https://doi.org/10.1007/JHEP02(2011)063} {\bibfield  {journal} {\bibinfo
  {journal} {JHEP}\ }\textbf {\bibinfo {volume} {2011}}\bibfield  {number}
  {\bibinfo  {number} { (02)},\ \bibinfo {pages} {063}},\ }\Eprint
  {https://arxiv.org/abs/1011.2440} {arXiv:1011.2440 [hep-th]} \BibitemShut
  {NoStop}%
\bibitem [{\citenamefont {Brandhuber}\ \emph {et~al.}(2011)\citenamefont
  {Brandhuber}, \citenamefont {Gurdogan}, \citenamefont {Mooney}, \citenamefont
  {Travaglini},\ and\ \citenamefont {Yang}}]{Brandhuber:2011}%
  \BibitemOpen
  \bibfield  {author} {\bibinfo {author} {\bibfnamefont {A.}~\bibnamefont
  {Brandhuber}}, \bibinfo {author} {\bibfnamefont {O.}~\bibnamefont
  {Gurdogan}}, \bibinfo {author} {\bibfnamefont {R.}~\bibnamefont {Mooney}},
  \bibinfo {author} {\bibfnamefont {G.}~\bibnamefont {Travaglini}},\ and\
  \bibinfo {author} {\bibfnamefont {G.}~\bibnamefont {Yang}},\ }\bibfield
  {title} {\bibinfo {title} {{Harmony of Super Form Factors}},\ }\href
  {https://doi.org/10.1007/JHEP10(2011)046} {\bibfield  {journal} {\bibinfo
  {journal} {JHEP}\ }\textbf {\bibinfo {volume} {2011}}\bibfield  {number}
  {\bibinfo  {number} { (10)},\ \bibinfo {pages} {046}},\ }\Eprint
  {https://arxiv.org/abs/1107.5067} {arXiv:1107.5067 [hep-th]} \BibitemShut
  {NoStop}%
\bibitem [{\citenamefont {Bork}(2013)}]{Bork:2013}%
  \BibitemOpen
  \bibfield  {author} {\bibinfo {author} {\bibfnamefont {L.~V.}\ \bibnamefont
  {Bork}},\ }\bibfield  {title} {\bibinfo {title} {{On NMHV form factors in N=4
  SYM theory from generalized unitarity}},\ }\href
  {https://doi.org/10.1007/JHEP01(2013)049} {\bibfield  {journal} {\bibinfo
  {journal} {JHEP}\ }\textbf {\bibinfo {volume} {2013}}\bibfield  {number}
  {\bibinfo  {number} { (01)},\ \bibinfo {pages} {049}},\ }\Eprint
  {https://arxiv.org/abs/1203.2596} {arXiv:1203.2596 [hep-th]} \BibitemShut
  {NoStop}%
\bibitem [{\citenamefont {Frassek}\ \emph {et~al.}(2016)\citenamefont
  {Frassek}, \citenamefont {Meidinger}, \citenamefont {Nandan},\ and\
  \citenamefont {Wilhelm}}]{Frassek:2016}%
  \BibitemOpen
  \bibfield  {author} {\bibinfo {author} {\bibfnamefont {R.}~\bibnamefont
  {Frassek}}, \bibinfo {author} {\bibfnamefont {D.}~\bibnamefont {Meidinger}},
  \bibinfo {author} {\bibfnamefont {D.}~\bibnamefont {Nandan}},\ and\ \bibinfo
  {author} {\bibfnamefont {M.}~\bibnamefont {Wilhelm}},\ }\bibfield  {title}
  {\bibinfo {title} {{On-shell diagrams, Gra{\ss}mannians and integrability for
  form factors}},\ }\href {https://doi.org/10.1007/JHEP01(2016)182} {\bibfield
  {journal} {\bibinfo  {journal} {JHEP}\ }\textbf {\bibinfo {volume}
  {2016}}\bibfield  {number} {\bibinfo  {number} { (01)},\ \bibinfo {pages}
  {182}},\ }\Eprint {https://arxiv.org/abs/1506.08192} {arXiv:1506.08192
  [hep-th]} \BibitemShut {NoStop}%
\bibitem [{\citenamefont {Dixon}\ \emph {et~al.}(2022)\citenamefont {Dixon},
  \citenamefont {Gurdogan}, \citenamefont {McLeod},\ and\ \citenamefont
  {Wilhelm}}]{DixonEtAl:2022EightLoops}%
  \BibitemOpen
  \bibfield  {author} {\bibinfo {author} {\bibfnamefont {L.~J.}\ \bibnamefont
  {Dixon}}, \bibinfo {author} {\bibfnamefont {O.}~\bibnamefont {Gurdogan}},
  \bibinfo {author} {\bibfnamefont {A.~J.}\ \bibnamefont {McLeod}},\ and\
  \bibinfo {author} {\bibfnamefont {M.}~\bibnamefont {Wilhelm}},\ }\bibfield
  {title} {\bibinfo {title} {{Bootstrapping a stress-tensor form factor through
  eight loops}},\ }\href {https://doi.org/10.1007/JHEP07(2022)153} {\bibfield
  {journal} {\bibinfo  {journal} {JHEP}\ }\textbf {\bibinfo {volume}
  {2022}}\bibfield  {number} {\bibinfo  {number} { (07)},\ \bibinfo {pages}
  {153}},\ }\Eprint {https://arxiv.org/abs/2204.11901} {arXiv:2204.11901
  [hep-th]} \BibitemShut {NoStop}%
\bibitem [{\citenamefont {Li}(2025)}]{Li:2025PeriodicWilsonLoop}%
  \BibitemOpen
  \bibfield  {author} {\bibinfo {author} {\bibfnamefont {Z.}~\bibnamefont
  {Li}},\ }\bibfield  {title} {\bibinfo {title} {{Two-loop MHV form factors
  from the periodic Wilson loop}},\ }\href
  {https://doi.org/10.1007/JHEP05(2025)209} {\bibfield  {journal} {\bibinfo
  {journal} {JHEP}\ }\textbf {\bibinfo {volume} {2025}}\bibfield  {number}
  {\bibinfo  {number} { (05)},\ \bibinfo {pages} {209}},\ }\Eprint
  {https://arxiv.org/abs/2412.17974} {arXiv:2412.17974 [hep-th]} \BibitemShut
  {NoStop}%
\bibitem [{\citenamefont {Dixon}\ \emph {et~al.}(2023)\citenamefont {Dixon},
  \citenamefont {G{\"u}rdo{\u{g}}an}, \citenamefont {Liu}, \citenamefont
  {McLeod},\ and\ \citenamefont {Wilhelm}}]{DixonSelfDual:2023}%
  \BibitemOpen
  \bibfield  {author} {\bibinfo {author} {\bibfnamefont {L.~J.}\ \bibnamefont
  {Dixon}}, \bibinfo {author} {\bibfnamefont {{\"O}.}~\bibnamefont
  {G{\"u}rdo{\u{g}}an}}, \bibinfo {author} {\bibfnamefont {Y.-T.}\ \bibnamefont
  {Liu}}, \bibinfo {author} {\bibfnamefont {A.~J.}\ \bibnamefont {McLeod}},\
  and\ \bibinfo {author} {\bibfnamefont {M.}~\bibnamefont {Wilhelm}},\
  }\bibfield  {title} {\bibinfo {title} {{Antipodal Self-Duality for a
  Four-Particle Form Factor}},\ }\href
  {https://doi.org/10.1103/PhysRevLett.130.111601} {\bibfield  {journal}
  {\bibinfo  {journal} {Phys. Rev. Lett.}\ }\textbf {\bibinfo {volume} {130}},\
  \bibinfo {pages} {111601} (\bibinfo {year} {2023})},\ \Eprint
  {https://arxiv.org/abs/2212.02410} {arXiv:2212.02410 [hep-th]} \BibitemShut
  {NoStop}%
\bibitem [{\citenamefont {Chicherin}\ \emph {et~al.}(2021)\citenamefont
  {Chicherin}, \citenamefont {Henn},\ and\ \citenamefont
  {Papathanasiou}}]{ChicherinHennPapathanasiou:2021}%
  \BibitemOpen
  \bibfield  {author} {\bibinfo {author} {\bibfnamefont {D.}~\bibnamefont
  {Chicherin}}, \bibinfo {author} {\bibfnamefont {J.~M.}\ \bibnamefont
  {Henn}},\ and\ \bibinfo {author} {\bibfnamefont {G.}~\bibnamefont
  {Papathanasiou}},\ }\bibfield  {title} {\bibinfo {title} {{Cluster algebras
  for Feynman integrals}},\ }\href
  {https://doi.org/10.1103/PhysRevLett.126.091603} {\bibfield  {journal}
  {\bibinfo  {journal} {Phys. Rev. Lett.}\ }\textbf {\bibinfo {volume} {126}},\
  \bibinfo {pages} {091603} (\bibinfo {year} {2021})},\ \Eprint
  {https://arxiv.org/abs/2012.12285} {arXiv:2012.12285 [hep-th]} \BibitemShut
  {NoStop}%
\bibitem [{\citenamefont {He}\ and\ \citenamefont
  {Liu}(2026)}]{HeLiuCluster:2026}%
  \BibitemOpen
  \bibfield  {author} {\bibinfo {author} {\bibfnamefont {S.}~\bibnamefont
  {He}}\ and\ \bibinfo {author} {\bibfnamefont {J.}~\bibnamefont {Liu}},\
  }\href@noop {} {\bibinfo {title} {{Form Factor Alphabets and Antipodal
  Duality from Cluster Algebras}}} (\bibinfo {year} {2026}),\ \Eprint
  {https://arxiv.org/abs/2609.04839} {arXiv:2609.04839 [hep-th]} \BibitemShut
  {NoStop}%
\bibitem [{\citenamefont {Guo}\ \emph {et~al.}(2025)\citenamefont {Guo},
  \citenamefont {Wang}, \citenamefont {Yang},\ and\ \citenamefont
  {Yin}}]{GuoEtAl:2025MHV}%
  \BibitemOpen
  \bibfield  {author} {\bibinfo {author} {\bibfnamefont {Y.}~\bibnamefont
  {Guo}}, \bibinfo {author} {\bibfnamefont {L.}~\bibnamefont {Wang}}, \bibinfo
  {author} {\bibfnamefont {G.}~\bibnamefont {Yang}},\ and\ \bibinfo {author}
  {\bibfnamefont {Y.}~\bibnamefont {Yin}},\ }\bibfield  {title} {\bibinfo
  {title} {{Analytic two-loop four-point form factor of the stress-tensor
  supermultiplet in $ \mathcal{N} $ = 4 SYM}},\ }\href
  {https://doi.org/10.1007/JHEP02(2025)002} {\bibfield  {journal} {\bibinfo
  {journal} {JHEP}\ }\textbf {\bibinfo {volume} {2025}}\bibfield  {number}
  {\bibinfo  {number} { (02)},\ \bibinfo {pages} {002}},\ }\Eprint
  {https://arxiv.org/abs/2409.12445} {arXiv:2409.12445 [hep-th]} \BibitemShut
  {NoStop}%
\bibitem [{\citenamefont {Dixon}\ and\ \citenamefont
  {Xin}(2025)}]{DixonXin:2025}%
  \BibitemOpen
  \bibfield  {author} {\bibinfo {author} {\bibfnamefont {L.~J.}\ \bibnamefont
  {Dixon}}\ and\ \bibinfo {author} {\bibfnamefont {S.}~\bibnamefont {Xin}},\
  }\bibfield  {title} {\bibinfo {title} {{A two-loop four-point form factor at
  function level}},\ }\href {https://doi.org/10.1007/JHEP01(2025)012}
  {\bibfield  {journal} {\bibinfo  {journal} {JHEP}\ }\textbf {\bibinfo
  {volume} {2025}}\bibfield  {number} {\bibinfo  {number} { (01)},\ \bibinfo
  {pages} {012}},\ }\bibinfo {note} {[Erratum: JHEP 03, 159 (2026)]},\ \Eprint
  {https://arxiv.org/abs/2411.01571} {arXiv:2411.01571 [hep-th]} \BibitemShut
  {NoStop}%
\bibitem [{\citenamefont {He}\ and\ \citenamefont
  {Jiang}(2026)}]{HeJiang:2026}%
  \BibitemOpen
  \bibfield  {author} {\bibinfo {author} {\bibfnamefont {S.}~\bibnamefont
  {He}}\ and\ \bibinfo {author} {\bibfnamefont {X.}~\bibnamefont {Jiang}},\
  }\href@noop {} {\bibinfo {title} {{A full-color four-point form factor}}}
  (\bibinfo {year} {2026}),\ \Eprint {https://arxiv.org/abs/2611.25591}
  {arXiv:2611.25591 [hep-th]} \BibitemShut {NoStop}%
\bibitem [{\citenamefont {Cai}\ \emph {et~al.}(2026)\citenamefont {Cai},
  \citenamefont {Guo},\ and\ \citenamefont {Yang}}]{CaiGuoYang:2026}%
  \BibitemOpen
  \bibfield  {author} {\bibinfo {author} {\bibfnamefont {Z.}~\bibnamefont
  {Cai}}, \bibinfo {author} {\bibfnamefont {Y.}~\bibnamefont {Guo}},\ and\
  \bibinfo {author} {\bibfnamefont {G.}~\bibnamefont {Yang}},\ }\href@noop {}
  {\bibinfo {title} {{Full-Color Two-loop Four-point Form Factor in N=4 SYM
  with Global CK-duality}}} (\bibinfo {year} {2026}),\ \Eprint
  {https://arxiv.org/abs/2611.26194} {arXiv:2611.26194 [hep-th]} \BibitemShut
  {NoStop}%
\bibitem [{\citenamefont {He}\ \emph {et~al.}(2026)\citenamefont {He},
  \citenamefont {Liu},\ and\ \citenamefont {Yang}}]{HeLiuYang:2026}%
  \BibitemOpen
  \bibfield  {author} {\bibinfo {author} {\bibfnamefont {S.}~\bibnamefont
  {He}}, \bibinfo {author} {\bibfnamefont {J.}~\bibnamefont {Liu}},\ and\
  \bibinfo {author} {\bibfnamefont {Q.}~\bibnamefont {Yang}},\ }\href@noop {}
  {\bibinfo {title} {{Bootstrapping the Four-Point NMHV Stress-Tensor Form
  Factor}}} (\bibinfo {year} {2026}),\ \Eprint
  {https://arxiv.org/abs/2605.28955} {arXiv:2605.28955 [hep-th]} \BibitemShut
  {NoStop}%
\bibitem [{\citenamefont {Chicherin}\ \emph {et~al.}(2022)\citenamefont
  {Chicherin}, \citenamefont {Sotnikov},\ and\ \citenamefont
  {Zoia}}]{Chicherin:2022}%
  \BibitemOpen
  \bibfield  {author} {\bibinfo {author} {\bibfnamefont {D.}~\bibnamefont
  {Chicherin}}, \bibinfo {author} {\bibfnamefont {V.}~\bibnamefont
  {Sotnikov}},\ and\ \bibinfo {author} {\bibfnamefont {S.}~\bibnamefont
  {Zoia}},\ }\bibfield  {title} {\bibinfo {title} {{Pentagon functions for
  one-mass planar scattering amplitudes}},\ }\href
  {https://doi.org/10.1007/JHEP01(2022)096} {\bibfield  {journal} {\bibinfo
  {journal} {JHEP}\ }\textbf {\bibinfo {volume} {2022}}\bibfield  {number}
  {\bibinfo  {number} { (01)},\ \bibinfo {pages} {096}},\ }\Eprint
  {https://arxiv.org/abs/2110.10111} {arXiv:2110.10111 [hep-ph]} \BibitemShut
  {NoStop}%
\bibitem [{\citenamefont {Abreu}\ \emph {et~al.}(2024)\citenamefont {Abreu},
  \citenamefont {Chicherin}, \citenamefont {Ita}, \citenamefont {Page},
  \citenamefont {Sotnikov}, \citenamefont {Tschernow},\ and\ \citenamefont
  {Zoia}}]{Abreu:2024}%
  \BibitemOpen
  \bibfield  {author} {\bibinfo {author} {\bibfnamefont {S.}~\bibnamefont
  {Abreu}}, \bibinfo {author} {\bibfnamefont {D.}~\bibnamefont {Chicherin}},
  \bibinfo {author} {\bibfnamefont {H.}~\bibnamefont {Ita}}, \bibinfo {author}
  {\bibfnamefont {B.}~\bibnamefont {Page}}, \bibinfo {author} {\bibfnamefont
  {V.}~\bibnamefont {Sotnikov}}, \bibinfo {author} {\bibfnamefont
  {W.}~\bibnamefont {Tschernow}},\ and\ \bibinfo {author} {\bibfnamefont
  {S.}~\bibnamefont {Zoia}},\ }\bibfield  {title} {\bibinfo {title} {{All
  Two-Loop Feynman Integrals for Five-Point One-Mass Scattering}},\ }\href
  {https://doi.org/10.1103/PhysRevLett.132.141601} {\bibfield  {journal}
  {\bibinfo  {journal} {Phys. Rev. Lett.}\ }\textbf {\bibinfo {volume} {132}},\
  \bibinfo {pages} {141601} (\bibinfo {year} {2024})},\ \Eprint
  {https://arxiv.org/abs/2306.15431} {arXiv:2306.15431 [hep-ph]} \BibitemShut
  {NoStop}%
\bibitem [{\citenamefont {Becher}\ and\ \citenamefont
  {Neubert}(2009)}]{Becher:2009}%
  \BibitemOpen
  \bibfield  {author} {\bibinfo {author} {\bibfnamefont {T.}~\bibnamefont
  {Becher}}\ and\ \bibinfo {author} {\bibfnamefont {M.}~\bibnamefont
  {Neubert}},\ }\bibfield  {title} {\bibinfo {title} {{Infrared singularities
  of scattering amplitudes in perturbative QCD}},\ }\href
  {https://doi.org/10.1103/PhysRevLett.102.162001} {\bibfield  {journal}
  {\bibinfo  {journal} {Phys. Rev. Lett.}\ }\textbf {\bibinfo {volume} {102}},\
  \bibinfo {pages} {162001} (\bibinfo {year} {2009})},\ \bibinfo {note}
  {[Erratum: Phys.Rev.Lett. 111, 199905 (2013)]},\ \Eprint
  {https://arxiv.org/abs/0901.0722} {arXiv:0901.0722 [hep-ph]} \BibitemShut
  {NoStop}%
\bibitem [{\citenamefont {Gardi}\ and\ \citenamefont
  {Magnea}(2009)}]{Gardi:2009}%
  \BibitemOpen
  \bibfield  {author} {\bibinfo {author} {\bibfnamefont {E.}~\bibnamefont
  {Gardi}}\ and\ \bibinfo {author} {\bibfnamefont {L.}~\bibnamefont {Magnea}},\
  }\bibfield  {title} {\bibinfo {title} {{Factorization constraints for soft
  anomalous dimensions in QCD scattering amplitudes}},\ }\href
  {https://doi.org/10.1088/1126-6708/2009/03/079} {\bibfield  {journal}
  {\bibinfo  {journal} {JHEP}\ }\textbf {\bibinfo {volume} {2009}}\bibfield
  {number} {\bibinfo  {number} { (03)},\ \bibinfo {pages} {079}},\ }\Eprint
  {https://arxiv.org/abs/0901.1091} {arXiv:0901.1091 [hep-ph]} \BibitemShut
  {NoStop}%
\bibitem [{\citenamefont {Naculich}(2012)}]{Naculich:2012}%
  \BibitemOpen
  \bibfield  {author} {\bibinfo {author} {\bibfnamefont {S.~G.}\ \bibnamefont
  {Naculich}},\ }\bibfield  {title} {\bibinfo {title} {{All-loop group-theory
  constraints for color-ordered SU(N) gauge-theory amplitudes}},\ }\href
  {https://doi.org/10.1016/j.physletb.2011.12.010} {\bibfield  {journal}
  {\bibinfo  {journal} {Phys. Lett. B}\ }\textbf {\bibinfo {volume} {707}},\
  \bibinfo {pages} {191} (\bibinfo {year} {2012})},\ \Eprint
  {https://arxiv.org/abs/1110.1859} {arXiv:1110.1859 [hep-th]} \BibitemShut
  {NoStop}%
\bibitem [{\citenamefont {Naculich}\ and\ \citenamefont
  {Osathapan}(2024)}]{NaculichOsathapan:2024}%
  \BibitemOpen
  \bibfield  {author} {\bibinfo {author} {\bibfnamefont {S.~G.}\ \bibnamefont
  {Naculich}}\ and\ \bibinfo {author} {\bibfnamefont {A.}~\bibnamefont
  {Osathapan}},\ }\bibfield  {title} {\bibinfo {title} {{All-loop group-theory
  constraints for four-point amplitudes of SU(N), SO(N), and Sp(N) gauge
  theories}},\ }\href {https://doi.org/10.1007/JHEP10(2024)221} {\bibfield
  {journal} {\bibinfo  {journal} {JHEP}\ }\textbf {\bibinfo {volume}
  {2024}}\bibfield  {number} {\bibinfo  {number} { (10)},\ \bibinfo {pages}
  {221}},\ }\Eprint {https://arxiv.org/abs/2407.03403} {arXiv:2407.03403
  [hep-th]} \BibitemShut {NoStop}%
\bibitem [{\citenamefont {Bork}(2014)}]{Bork:2014}%
  \BibitemOpen
  \bibfield  {author} {\bibinfo {author} {\bibfnamefont {L.~V.}\ \bibnamefont
  {Bork}},\ }\bibfield  {title} {\bibinfo {title} {{On form factors in $
  \mathcal{N}=4 $ SYM theory and polytopes}},\ }\href
  {https://doi.org/10.1007/JHEP12(2014)111} {\bibfield  {journal} {\bibinfo
  {journal} {JHEP}\ }\textbf {\bibinfo {volume} {2014}}\bibfield  {number}
  {\bibinfo  {number} { (12)},\ \bibinfo {pages} {111}},\ }\Eprint
  {https://arxiv.org/abs/1407.5568} {arXiv:1407.5568 [hep-th]} \BibitemShut
  {NoStop}%
\bibitem [{\citenamefont {Koster}\ \emph {et~al.}(2017)\citenamefont {Koster},
  \citenamefont {Mitev}, \citenamefont {Staudacher},\ and\ \citenamefont
  {Wilhelm}}]{KosterEtAl:2017}%
  \BibitemOpen
  \bibfield  {author} {\bibinfo {author} {\bibfnamefont {L.}~\bibnamefont
  {Koster}}, \bibinfo {author} {\bibfnamefont {V.}~\bibnamefont {Mitev}},
  \bibinfo {author} {\bibfnamefont {M.}~\bibnamefont {Staudacher}},\ and\
  \bibinfo {author} {\bibfnamefont {M.}~\bibnamefont {Wilhelm}},\ }\bibfield
  {title} {\bibinfo {title} {{On Form Factors and Correlation Functions in
  Twistor Space}},\ }\href {https://doi.org/10.1007/JHEP03(2017)131} {\bibfield
   {journal} {\bibinfo  {journal} {JHEP}\ }\textbf {\bibinfo {volume}
  {2017}}\bibfield  {number} {\bibinfo  {number} { (03)},\ \bibinfo {pages}
  {131}},\ }\Eprint {https://arxiv.org/abs/1611.08599} {arXiv:1611.08599
  [hep-th]} \BibitemShut {NoStop}%
\bibitem [{\citenamefont {Bianchi}\ \emph {et~al.}(2019)\citenamefont
  {Bianchi}, \citenamefont {Brandhuber}, \citenamefont {Panerai},\ and\
  \citenamefont {Travaglini}}]{BianchiEtAl:2019}%
  \BibitemOpen
  \bibfield  {author} {\bibinfo {author} {\bibfnamefont {L.}~\bibnamefont
  {Bianchi}}, \bibinfo {author} {\bibfnamefont {A.}~\bibnamefont {Brandhuber}},
  \bibinfo {author} {\bibfnamefont {R.}~\bibnamefont {Panerai}},\ and\ \bibinfo
  {author} {\bibfnamefont {G.}~\bibnamefont {Travaglini}},\ }\bibfield  {title}
  {\bibinfo {title} {{Dual conformal invariance for form factors}},\ }\href
  {https://doi.org/10.1007/JHEP02(2019)134} {\bibfield  {journal} {\bibinfo
  {journal} {JHEP}\ }\textbf {\bibinfo {volume} {2019}}\bibfield  {number}
  {\bibinfo  {number} { (02)},\ \bibinfo {pages} {134}},\ }\Eprint
  {https://arxiv.org/abs/1812.10468} {arXiv:1812.10468 [hep-th]} \BibitemShut
  {NoStop}%
\bibitem [{\citenamefont {Abreu}\ \emph {et~al.}(2023)\citenamefont {Abreu},
  \citenamefont {Chicherin}, \citenamefont {Ita}, \citenamefont {Page},
  \citenamefont {Sotnikov}, \citenamefont {Tschernow},\ and\ \citenamefont
  {Zoia}}]{Abreu:2023data}%
  \BibitemOpen
  \bibfield  {author} {\bibinfo {author} {\bibfnamefont {S.}~\bibnamefont
  {Abreu}}, \bibinfo {author} {\bibfnamefont {D.}~\bibnamefont {Chicherin}},
  \bibinfo {author} {\bibfnamefont {H.}~\bibnamefont {Ita}}, \bibinfo {author}
  {\bibfnamefont {B.}~\bibnamefont {Page}}, \bibinfo {author} {\bibfnamefont
  {V.}~\bibnamefont {Sotnikov}}, \bibinfo {author} {\bibfnamefont
  {W.}~\bibnamefont {Tschernow}},\ and\ \bibinfo {author} {\bibfnamefont
  {S.}~\bibnamefont {Zoia}},\ }\href {https://doi.org/10.5281/zenodo.8085740}
  {\bibinfo {title} {Supplementary material for {All Two-Loop Feynman Integrals
  for Five-Point One-Mass Scattering}}},\ \bibinfo {howpublished} {Zenodo}
  (\bibinfo {year} {2023}),\ \bibinfo {note} {version 1.1}\BibitemShut
  {NoStop}%
\bibitem [{\citenamefont {Abreu}\ \emph {et~al.}(2020)\citenamefont {Abreu},
  \citenamefont {Ita}, \citenamefont {Moriello}, \citenamefont {Page},
  \citenamefont {Tschernow},\ and\ \citenamefont
  {Zeng}}]{AbreuItaMoriello:2020}%
  \BibitemOpen
  \bibfield  {author} {\bibinfo {author} {\bibfnamefont {S.}~\bibnamefont
  {Abreu}}, \bibinfo {author} {\bibfnamefont {H.}~\bibnamefont {Ita}}, \bibinfo
  {author} {\bibfnamefont {F.}~\bibnamefont {Moriello}}, \bibinfo {author}
  {\bibfnamefont {B.}~\bibnamefont {Page}}, \bibinfo {author} {\bibfnamefont
  {W.}~\bibnamefont {Tschernow}},\ and\ \bibinfo {author} {\bibfnamefont
  {M.}~\bibnamefont {Zeng}},\ }\bibfield  {title} {\bibinfo {title} {{Two-Loop
  Integrals for Planar Five-Point One-Mass Processes}},\ }\href
  {https://doi.org/10.1007/JHEP11(2020)117} {\bibfield  {journal} {\bibinfo
  {journal} {JHEP}\ }\textbf {\bibinfo {volume} {2020}}\bibfield  {number}
  {\bibinfo  {number} { (11)},\ \bibinfo {pages} {117}},\ }\Eprint
  {https://arxiv.org/abs/2005.04195} {arXiv:2005.04195 [hep-ph]} \BibitemShut
  {NoStop}%
\bibitem [{\citenamefont {Steinmann}(1960{\natexlab{a}})}]{Steinmann:1960a}%
  \BibitemOpen
  \bibfield  {author} {\bibinfo {author} {\bibfnamefont {O.}~\bibnamefont
  {Steinmann}},\ }\bibfield  {title} {\bibinfo {title} {{{\"U}ber den
  Zusammenhang zwischen den Wightmanfunktionen und den retardierten
  Kommutatoren}},\ }\href {https://doi.org/10.5169/seals-113076} {\bibfield
  {journal} {\bibinfo  {journal} {Helv. Phys. Acta}\ }\textbf {\bibinfo
  {volume} {33}},\ \bibinfo {pages} {257} (\bibinfo {year}
  {1960}{\natexlab{a}})}\BibitemShut {NoStop}%
\bibitem [{\citenamefont {Steinmann}(1960{\natexlab{b}})}]{Steinmann:1960b}%
  \BibitemOpen
  \bibfield  {author} {\bibinfo {author} {\bibfnamefont {O.}~\bibnamefont
  {Steinmann}},\ }\bibfield  {title} {\bibinfo {title} {{Wightman-Funktionen
  und retardierte Kommutatoren. II}},\ }\href
  {https://doi.org/10.5169/seals-113079} {\bibfield  {journal} {\bibinfo
  {journal} {Helv. Phys. Acta}\ }\textbf {\bibinfo {volume} {33}},\ \bibinfo
  {pages} {347} (\bibinfo {year} {1960}{\natexlab{b}})}\BibitemShut {NoStop}%
\bibitem [{\citenamefont {Caron-Huot}\ \emph {et~al.}(2016)\citenamefont
  {Caron-Huot}, \citenamefont {Dixon}, \citenamefont {McLeod},\ and\
  \citenamefont {von Hippel}}]{CaronHuot:2016}%
  \BibitemOpen
  \bibfield  {author} {\bibinfo {author} {\bibfnamefont {S.}~\bibnamefont
  {Caron-Huot}}, \bibinfo {author} {\bibfnamefont {L.~J.}\ \bibnamefont
  {Dixon}}, \bibinfo {author} {\bibfnamefont {A.}~\bibnamefont {McLeod}},\ and\
  \bibinfo {author} {\bibfnamefont {M.}~\bibnamefont {von Hippel}},\ }\bibfield
   {title} {\bibinfo {title} {{Bootstrapping a Five-Loop Amplitude Using
  Steinmann Relations}},\ }\href
  {https://doi.org/10.1103/PhysRevLett.117.241601} {\bibfield  {journal}
  {\bibinfo  {journal} {Phys. Rev. Lett.}\ }\textbf {\bibinfo {volume} {117}},\
  \bibinfo {pages} {241601} (\bibinfo {year} {2016})},\ \Eprint
  {https://arxiv.org/abs/1609.00669} {arXiv:1609.00669 [hep-th]} \BibitemShut
  {NoStop}%
\bibitem [{\citenamefont {Brandhuber}\ \emph {et~al.}(2012)\citenamefont
  {Brandhuber}, \citenamefont {Travaglini},\ and\ \citenamefont
  {Yang}}]{Brandhuber:2012}%
  \BibitemOpen
  \bibfield  {author} {\bibinfo {author} {\bibfnamefont {A.}~\bibnamefont
  {Brandhuber}}, \bibinfo {author} {\bibfnamefont {G.}~\bibnamefont
  {Travaglini}},\ and\ \bibinfo {author} {\bibfnamefont {G.}~\bibnamefont
  {Yang}},\ }\bibfield  {title} {\bibinfo {title} {{Analytic two-loop form
  factors in N=4 SYM}},\ }\href {https://doi.org/10.1007/JHEP05(2012)082}
  {\bibfield  {journal} {\bibinfo  {journal} {JHEP}\ }\textbf {\bibinfo
  {volume} {2012}}\bibfield  {number} {\bibinfo  {number} { (05)},\ \bibinfo
  {pages} {082}},\ }\Eprint {https://arxiv.org/abs/1201.4170} {arXiv:1201.4170
  [hep-th]} \BibitemShut {NoStop}%
\bibitem [{\citenamefont {Bern}\ \emph {et~al.}(2004)\citenamefont {Bern},
  \citenamefont {Dixon},\ and\ \citenamefont
  {Kosower}}]{BernDixonKosower:2004}%
  \BibitemOpen
  \bibfield  {author} {\bibinfo {author} {\bibfnamefont {Z.}~\bibnamefont
  {Bern}}, \bibinfo {author} {\bibfnamefont {L.~J.}\ \bibnamefont {Dixon}},\
  and\ \bibinfo {author} {\bibfnamefont {D.~A.}\ \bibnamefont {Kosower}},\
  }\bibfield  {title} {\bibinfo {title} {{Two-loop g ---{\ensuremath{>}} gg
  splitting amplitudes in QCD}},\ }\href
  {https://doi.org/10.1088/1126-6708/2004/08/012} {\bibfield  {journal}
  {\bibinfo  {journal} {JHEP}\ }\textbf {\bibinfo {volume} {2004}}\bibfield
  {number} {\bibinfo  {number} { (08)},\ \bibinfo {pages} {012}},\ }\Eprint
  {https://arxiv.org/abs/hep-ph/0404293} {arXiv:hep-ph/0404293} \BibitemShut
  {NoStop}%
\bibitem [{\citenamefont {Dixon}\ \emph {et~al.}(2020)\citenamefont {Dixon},
  \citenamefont {Herrmann}, \citenamefont {Yan},\ and\ \citenamefont
  {Zhu}}]{DixonSoft:2020}%
  \BibitemOpen
  \bibfield  {author} {\bibinfo {author} {\bibfnamefont {L.~J.}\ \bibnamefont
  {Dixon}}, \bibinfo {author} {\bibfnamefont {E.}~\bibnamefont {Herrmann}},
  \bibinfo {author} {\bibfnamefont {K.}~\bibnamefont {Yan}},\ and\ \bibinfo
  {author} {\bibfnamefont {H.~X.}\ \bibnamefont {Zhu}},\ }\bibfield  {title}
  {\bibinfo {title} {{Soft gluon emission at two loops in full color}},\ }\href
  {https://doi.org/10.1007/JHEP05(2020)135} {\bibfield  {journal} {\bibinfo
  {journal} {JHEP}\ }\textbf {\bibinfo {volume} {2020}}\bibfield  {number}
  {\bibinfo  {number} { (05)},\ \bibinfo {pages} {135}},\ }\bibinfo {note}
  {[Erratum: JHEP 06, 143 (2024)]},\ \Eprint {https://arxiv.org/abs/1912.09370}
  {arXiv:1912.09370 [hep-ph]} \BibitemShut {NoStop}%
\bibitem [{\citenamefont {Bern}\ \emph {et~al.}(1994)\citenamefont {Bern},
  \citenamefont {Dixon}, \citenamefont {Dunbar},\ and\ \citenamefont
  {Kosower}}]{Bern:1994unitarity}%
  \BibitemOpen
  \bibfield  {author} {\bibinfo {author} {\bibfnamefont {Z.}~\bibnamefont
  {Bern}}, \bibinfo {author} {\bibfnamefont {L.~J.}\ \bibnamefont {Dixon}},
  \bibinfo {author} {\bibfnamefont {D.~C.}\ \bibnamefont {Dunbar}},\ and\
  \bibinfo {author} {\bibfnamefont {D.~A.}\ \bibnamefont {Kosower}},\
  }\bibfield  {title} {\bibinfo {title} {{One loop n point gauge theory
  amplitudes, unitarity and collinear limits}},\ }\href
  {https://doi.org/10.1016/0550-3213(94)90179-1} {\bibfield  {journal}
  {\bibinfo  {journal} {Nucl. Phys. B}\ }\textbf {\bibinfo {volume} {425}},\
  \bibinfo {pages} {217} (\bibinfo {year} {1994})},\ \Eprint
  {https://arxiv.org/abs/hep-ph/9403226} {arXiv:hep-ph/9403226} \BibitemShut
  {NoStop}%
\bibitem [{\citenamefont {Giele}\ \emph {et~al.}(2008)\citenamefont {Giele},
  \citenamefont {Kunszt},\ and\ \citenamefont {Melnikov}}]{Giele:2008}%
  \BibitemOpen
  \bibfield  {author} {\bibinfo {author} {\bibfnamefont {W.~T.}\ \bibnamefont
  {Giele}}, \bibinfo {author} {\bibfnamefont {Z.}~\bibnamefont {Kunszt}},\ and\
  \bibinfo {author} {\bibfnamefont {K.}~\bibnamefont {Melnikov}},\ }\bibfield
  {title} {\bibinfo {title} {{Full one-loop amplitudes from tree amplitudes}},\
  }\href {https://doi.org/10.1088/1126-6708/2008/04/049} {\bibfield  {journal}
  {\bibinfo  {journal} {JHEP}\ }\textbf {\bibinfo {volume} {2008}}\bibfield
  {number} {\bibinfo  {number} { (04)},\ \bibinfo {pages} {049}},\ }\Eprint
  {https://arxiv.org/abs/0801.2237} {arXiv:0801.2237 [hep-ph]} \BibitemShut
  {NoStop}%
\bibitem [{\citenamefont {Peraro}(2016)}]{Peraro:2016}%
  \BibitemOpen
  \bibfield  {author} {\bibinfo {author} {\bibfnamefont {T.}~\bibnamefont
  {Peraro}},\ }\bibfield  {title} {\bibinfo {title} {{Scattering amplitudes
  over finite fields and multivariate functional reconstruction}},\ }\href
  {https://doi.org/10.1007/JHEP12(2016)030} {\bibfield  {journal} {\bibinfo
  {journal} {JHEP}\ }\textbf {\bibinfo {volume} {2016}}\bibfield  {number}
  {\bibinfo  {number} { (12)},\ \bibinfo {pages} {030}},\ }\Eprint
  {https://arxiv.org/abs/1608.01902} {arXiv:1608.01902 [hep-ph]} \BibitemShut
  {NoStop}%
\bibitem [{\citenamefont {Abreu}\ \emph {et~al.}(2017)\citenamefont {Abreu},
  \citenamefont {Febres~Cordero}, \citenamefont {Ita}, \citenamefont {Jaquier},
  \citenamefont {Page},\ and\ \citenamefont {Zeng}}]{Abreu:2017unitarity}%
  \BibitemOpen
  \bibfield  {author} {\bibinfo {author} {\bibfnamefont {S.}~\bibnamefont
  {Abreu}}, \bibinfo {author} {\bibfnamefont {F.}~\bibnamefont
  {Febres~Cordero}}, \bibinfo {author} {\bibfnamefont {H.}~\bibnamefont {Ita}},
  \bibinfo {author} {\bibfnamefont {M.}~\bibnamefont {Jaquier}}, \bibinfo
  {author} {\bibfnamefont {B.}~\bibnamefont {Page}},\ and\ \bibinfo {author}
  {\bibfnamefont {M.}~\bibnamefont {Zeng}},\ }\bibfield  {title} {\bibinfo
  {title} {{Two-Loop Four-Gluon Amplitudes from Numerical Unitarity}},\ }\href
  {https://doi.org/10.1103/PhysRevLett.119.142001} {\bibfield  {journal}
  {\bibinfo  {journal} {Phys. Rev. Lett.}\ }\textbf {\bibinfo {volume} {119}},\
  \bibinfo {pages} {142001} (\bibinfo {year} {2017})},\ \Eprint
  {https://arxiv.org/abs/1703.05273} {arXiv:1703.05273 [hep-ph]} \BibitemShut
  {NoStop}%
\bibitem [{\citenamefont {Maierh{\"o}fer}\ \emph {et~al.}(2018)\citenamefont
  {Maierh{\"o}fer}, \citenamefont {Usovitsch},\ and\ \citenamefont
  {Uwer}}]{Maierhofer:2018}%
  \BibitemOpen
  \bibfield  {author} {\bibinfo {author} {\bibfnamefont {P.}~\bibnamefont
  {Maierh{\"o}fer}}, \bibinfo {author} {\bibfnamefont {J.}~\bibnamefont
  {Usovitsch}},\ and\ \bibinfo {author} {\bibfnamefont {P.}~\bibnamefont
  {Uwer}},\ }\bibfield  {title} {\bibinfo {title} {{Kira{\textemdash}A Feynman
  integral reduction program}},\ }\href
  {https://doi.org/10.1016/j.cpc.2018.04.012} {\bibfield  {journal} {\bibinfo
  {journal} {Comput. Phys. Commun.}\ }\textbf {\bibinfo {volume} {230}},\
  \bibinfo {pages} {99} (\bibinfo {year} {2018})},\ \Eprint
  {https://arxiv.org/abs/1705.05610} {arXiv:1705.05610 [hep-ph]} \BibitemShut
  {NoStop}%
\bibitem [{\citenamefont {Lange}\ \emph {et~al.}(2026)\citenamefont {Lange},
  \citenamefont {Usovitsch},\ and\ \citenamefont {Wu}}]{Lange:2026}%
  \BibitemOpen
  \bibfield  {author} {\bibinfo {author} {\bibfnamefont {F.}~\bibnamefont
  {Lange}}, \bibinfo {author} {\bibfnamefont {J.}~\bibnamefont {Usovitsch}},\
  and\ \bibinfo {author} {\bibfnamefont {Z.}~\bibnamefont {Wu}},\ }\bibfield
  {title} {\bibinfo {title} {{Kira 3: integral reduction with efficient seeding
  and optimized equation selection}},\ }\href
  {https://doi.org/10.1016/j.cpc.2025.109999} {\bibfield  {journal} {\bibinfo
  {journal} {Comput. Phys. Commun.}\ }\textbf {\bibinfo {volume} {322}},\
  \bibinfo {pages} {109999} (\bibinfo {year} {2026})},\ \Eprint
  {https://arxiv.org/abs/2505.20197} {arXiv:2505.20197 [hep-ph]} \BibitemShut
  {NoStop}%
\bibitem [{Note1()}]{Note1}%
  \BibitemOpen
  \bibinfo {note} {For the comparison we strip the planar powers of $N_c$ at
  each loop order and express the prefactors in a common convention. The
  conversion to the planar subtraction scheme adds $K\protect \mathcal
  H^{(1)}+\protect \tfrac 12 K^2\protect \mathcal H^{(0)}$ to our two-loop hard
  function at symbol level, where $K=\protect \tfrac 12\DOTSB \sum@ \slimits@
  _{i=1}^4\log ^2(s_{i,i+1}/q^2)$ with cyclic indices.}\BibitemShut {Stop}%
\bibitem [{\citenamefont {Caron-Huot}\ \emph {et~al.}(2019)\citenamefont
  {Caron-Huot}, \citenamefont {Dixon}, \citenamefont {Dulat}, \citenamefont
  {Von~Hippel}, \citenamefont {McLeod},\ and\ \citenamefont
  {Papathanasiou}}]{CaronHuot:2019}%
  \BibitemOpen
  \bibfield  {author} {\bibinfo {author} {\bibfnamefont {S.}~\bibnamefont
  {Caron-Huot}}, \bibinfo {author} {\bibfnamefont {L.~J.}\ \bibnamefont
  {Dixon}}, \bibinfo {author} {\bibfnamefont {F.}~\bibnamefont {Dulat}},
  \bibinfo {author} {\bibfnamefont {M.}~\bibnamefont {Von~Hippel}}, \bibinfo
  {author} {\bibfnamefont {A.~J.}\ \bibnamefont {McLeod}},\ and\ \bibinfo
  {author} {\bibfnamefont {G.}~\bibnamefont {Papathanasiou}},\ }\bibfield
  {title} {\bibinfo {title} {{The Cosmic Galois Group and Extended Steinmann
  Relations for Planar $\mathcal{N} = 4$ SYM Amplitudes}},\ }\href
  {https://doi.org/10.1007/JHEP09(2019)061} {\bibfield  {journal} {\bibinfo
  {journal} {JHEP}\ }\textbf {\bibinfo {volume} {2019}}\bibfield  {number}
  {\bibinfo  {number} { (09)},\ \bibinfo {pages} {061}},\ }\Eprint
  {https://arxiv.org/abs/1906.07116} {arXiv:1906.07116 [hep-th]} \BibitemShut
  {NoStop}%
\bibitem [{\citenamefont {Chen}\ \emph
  {et~al.}(2026{\natexlab{a}})\citenamefont {Chen}, \citenamefont {Chicherin},
  \citenamefont {Fox}, \citenamefont {Glover}, \citenamefont {Marcoli},
  \citenamefont {Sotnikov}, \citenamefont {Sun}, \citenamefont {Zhang},\ and\
  \citenamefont {Zoia}}]{ChenEtAl:2026FourJet}%
  \BibitemOpen
  \bibfield  {author} {\bibinfo {author} {\bibfnamefont {X.}~\bibnamefont
  {Chen}}, \bibinfo {author} {\bibfnamefont {D.}~\bibnamefont {Chicherin}},
  \bibinfo {author} {\bibfnamefont {E.}~\bibnamefont {Fox}}, \bibinfo {author}
  {\bibfnamefont {N.}~\bibnamefont {Glover}}, \bibinfo {author} {\bibfnamefont
  {M.}~\bibnamefont {Marcoli}}, \bibinfo {author} {\bibfnamefont
  {V.}~\bibnamefont {Sotnikov}}, \bibinfo {author} {\bibfnamefont
  {H.}~\bibnamefont {Sun}}, \bibinfo {author} {\bibfnamefont {H.}~\bibnamefont
  {Zhang}},\ and\ \bibinfo {author} {\bibfnamefont {S.}~\bibnamefont {Zoia}},\
  }\bibfield  {title} {\bibinfo {title} {{Four-Jet Rate in Electron-Positron
  Annihilation at Order $\alpha_s^4$}},\ }\href
  {https://doi.org/10.1103/xxg6-yd47} {\bibfield  {journal} {\bibinfo
  {journal} {Phys. Rev. Lett.}\ }\textbf {\bibinfo {volume} {136}},\ \bibinfo
  {pages} {251901} (\bibinfo {year} {2026}{\natexlab{a}})},\ \Eprint
  {https://arxiv.org/abs/2602.18185} {arXiv:2602.18185 [hep-ph]} \BibitemShut
  {NoStop}%
\bibitem [{\citenamefont {Chen}\ \emph
  {et~al.}(2026{\natexlab{b}})\citenamefont {Chen}, \citenamefont {Chicherin},
  \citenamefont {Fox}, \citenamefont {Glover}, \citenamefont {Marcoli},
  \citenamefont {Sotnikov}, \citenamefont {Sun}, \citenamefont {Zhang},\ and\
  \citenamefont {Zoia}}]{ChenEtAl:2026data}%
  \BibitemOpen
  \bibfield  {author} {\bibinfo {author} {\bibfnamefont {X.}~\bibnamefont
  {Chen}}, \bibinfo {author} {\bibfnamefont {D.}~\bibnamefont {Chicherin}},
  \bibinfo {author} {\bibfnamefont {E.}~\bibnamefont {Fox}}, \bibinfo {author}
  {\bibfnamefont {N.}~\bibnamefont {Glover}}, \bibinfo {author} {\bibfnamefont
  {M.}~\bibnamefont {Marcoli}}, \bibinfo {author} {\bibfnamefont
  {V.}~\bibnamefont {Sotnikov}}, \bibinfo {author} {\bibfnamefont
  {H.}~\bibnamefont {Sun}}, \bibinfo {author} {\bibfnamefont {H.}~\bibnamefont
  {Zhang}},\ and\ \bibinfo {author} {\bibfnamefont {S.}~\bibnamefont {Zoia}},\
  }\href {https://doi.org/10.5281/zenodo.18630975} {\bibinfo {title}
  {Supplementary material for {The Four-Jet Rate in Electron-Positron
  Annihilation at Order $\alpha_s^4$}}},\ \bibinfo {howpublished} {Zenodo}
  (\bibinfo {year} {2026}{\natexlab{b}}),\ \bibinfo {note} {version 1,
  decay-channel pentagon-function data}\BibitemShut {NoStop}%
\bibitem [{\citenamefont {Gehrmann}\ \emph
  {et~al.}(2012{\natexlab{a}})\citenamefont {Gehrmann}, \citenamefont {Henn},\
  and\ \citenamefont {Huber}}]{Gehrmann:2012Sudakov}%
  \BibitemOpen
  \bibfield  {author} {\bibinfo {author} {\bibfnamefont {T.}~\bibnamefont
  {Gehrmann}}, \bibinfo {author} {\bibfnamefont {J.~M.}\ \bibnamefont {Henn}},\
  and\ \bibinfo {author} {\bibfnamefont {T.}~\bibnamefont {Huber}},\ }\bibfield
   {title} {\bibinfo {title} {{The three-loop form factor in N=4 super
  Yang-Mills}},\ }\href {https://doi.org/10.1007/JHEP03(2012)101} {\bibfield
  {journal} {\bibinfo  {journal} {JHEP}\ }\textbf {\bibinfo {volume}
  {2012}}\bibfield  {number} {\bibinfo  {number} { (03)},\ \bibinfo {pages}
  {101}},\ }\Eprint {https://arxiv.org/abs/1112.4524} {arXiv:1112.4524
  [hep-th]} \BibitemShut {NoStop}%
\bibitem [{\citenamefont {Gehrmann}\ \emph
  {et~al.}(2012{\natexlab{b}})\citenamefont {Gehrmann}, \citenamefont
  {Jaquier}, \citenamefont {Glover},\ and\ \citenamefont
  {Koukoutsakis}}]{GehrmannEtAl:2012Higgs3Partons}%
  \BibitemOpen
  \bibfield  {author} {\bibinfo {author} {\bibfnamefont {T.}~\bibnamefont
  {Gehrmann}}, \bibinfo {author} {\bibfnamefont {M.}~\bibnamefont {Jaquier}},
  \bibinfo {author} {\bibfnamefont {E.~W.~N.}\ \bibnamefont {Glover}},\ and\
  \bibinfo {author} {\bibfnamefont {A.}~\bibnamefont {Koukoutsakis}},\
  }\bibfield  {title} {\bibinfo {title} {{Two-Loop QCD Corrections to the
  Helicity Amplitudes for $H \to$ 3 partons}},\ }\href
  {https://doi.org/10.1007/JHEP02(2012)056} {\bibfield  {journal} {\bibinfo
  {journal} {JHEP}\ }\textbf {\bibinfo {volume} {2012}}\bibfield  {number}
  {\bibinfo  {number} { (02)},\ \bibinfo {pages} {056}},\ }\Eprint
  {https://arxiv.org/abs/1112.3554} {arXiv:1112.3554 [hep-ph]} \BibitemShut
  {NoStop}%
\bibitem [{\citenamefont {De~Laurentis}\ \emph {et~al.}(2026)\citenamefont
  {De~Laurentis}, \citenamefont {Ita}, \citenamefont {Kuschke}, \citenamefont
  {Ruf},\ and\ \citenamefont {Sotnikov}}]{DeLaurentisEtAl:2026Higgs2Jets}%
  \BibitemOpen
  \bibfield  {author} {\bibinfo {author} {\bibfnamefont {G.}~\bibnamefont
  {De~Laurentis}}, \bibinfo {author} {\bibfnamefont {H.}~\bibnamefont {Ita}},
  \bibinfo {author} {\bibfnamefont {V.}~\bibnamefont {Kuschke}}, \bibinfo
  {author} {\bibfnamefont {M.}~\bibnamefont {Ruf}},\ and\ \bibinfo {author}
  {\bibfnamefont {V.}~\bibnamefont {Sotnikov}},\ }\bibfield  {title} {\bibinfo
  {title} {{Two-loop leading-color QCD corrections for Higgs plus two-jet
  production in the heavy-top limit}},\ }\href
  {https://doi.org/10.1007/JHEP07(2026)251} {\bibfield  {journal} {\bibinfo
  {journal} {JHEP}\ }\textbf {\bibinfo {volume} {2026}}\bibfield  {number}
  {\bibinfo  {number} { (07)},\ \bibinfo {pages} {251}},\ }\Eprint
  {https://arxiv.org/abs/2605.04009} {arXiv:2605.04009 [hep-ph]} \BibitemShut
  {NoStop}%
\end{thebibliography}%

\clearpage
\appendix
\onecolumngrid
\section{Loop normalization and infrared subtraction}
\label{app:subtraction}

In dimensional reduction we use the loop expansion, coupling,
and integration measure
\begin{equation}
\F=\sum_{n\geq0}a^n\F^{(n)},\qquad
a=\frac{g_{\rm YM}^2}{16\pi^2}(4\pi e^{-\gamma_E})^\eps,\qquad
\int_\ell=e^{\gamma_E\eps}\mu^{2\eps}
             \int\frac{\dd^{4-2\eps}\ell}{\ii\pi^{2-\eps}}.
\label{eq:measure}
\end{equation}

The color charge $\T_i$ on each external adjoint leg satisfies
$\T_i^2=N_c$.
For an $n$-point form factor with $n=2,3,4$, the dipole operator is
\begin{equation}
\M_n=-2\sum_{1\le i<j\le n}\T_i\cdot\T_j
             \log\frac{-s_{ij}-\ii0}{\mu^2}.
\label{eq:dipole}
\end{equation}
The two-loop dipole anomalous dimension
\cite{Becher:2009,Gardi:2009} gives the pure-pole operator $\Z_n$,
with the color-space identity implicit in the scalar terms,
\begin{equation}
\log\Z_n=a\left(-\frac{n\nc}{\eps^2}
       +\frac{\M_n}{\eps}\right)
   +a^2\left(\frac{n\nc^2\zeta_2}{2\eps^2}
       -\frac{\nc\zeta_2\M_n}{\eps}
       +\frac{n\nc^2\zeta_3}{2\eps}\right)+O(a^3).
\label{eq:Z}
\end{equation}
We write $\Z_n=1+a\Z_n^{(1)}+a^2\Z_n^{(2)}+O(a^3)$.

For the two-point form factor, color conservation gives $\M_2=2\nc L_q$,
where $L_q=\log[(-q^2-\ii0)/\mu^2]$.
In the same subtraction scheme, the two-point hard coefficients in
Eq.~\eqref{eq:boundary} are \cite{Gehrmann:2012Sudakov,Brandhuber:2012}
\begin{equation}
h_1=\zeta_2-L_q^2,\qquad
h_2=\tfrac12 h_1^2+2\zeta_2 L_q^2-2\zeta_3 L_q .
\label{eq:sudakov}
\end{equation}

\section{The full-color one-loop integrand}
\label{app:one-loop}

\begingroup
\setlength{\abovedisplayskip}{6pt plus 2pt minus 2pt}
\setlength{\belowdisplayskip}{6pt plus 2pt minus 2pt}
\setlength{\abovedisplayshortskip}{2pt plus 2pt}
\setlength{\belowdisplayshortskip}{4pt plus 2pt minus 2pt}

With the common delta functions stripped as in Sec.~\ref{sec:conventions},
the full-color one-loop NMHV form factor is the integral of
\begin{equation}
\mathcal I^{(1)}_{4,\mathrm{NMHV}}
=\sum_{i=1}^{3}\left(\nc S_i+\sum_{j=1}^{3}D_j\right)
  \mathcal I_{\sigma_i}
\label{eq:one-loop-color}
\end{equation}
with the measure in Eq.~\eqref{eq:measure}. We give $\mathcal I_{(1234)}$
below, and obtain the ordered integrands for $(1243)$ and $(1324)$ by
relabeling. We use $\mathcal D_{j_1\cdots j_k}=\prod_{m=1}^k\mathcal D_{j_m}$,
where $\mathcal D_j=(\ell-K_j)^2+\ii0$ with
$K_j=\sum_{n=0}^{j-1}p_{r+n}$ for $j=1,\ldots,4$ and $K_0=0$. We write
$\mu_\ell^2=\bar\ell^2-\ell^2$, with $\bar\ell$ the four-dimensional
part of $\ell$. In terms of the $R$-invariants and triangle prefactors
of Sec.~\ref{sec:bootstrap}, the ordered integrand is
\begin{equation}
\begin{aligned}
\mathcal I_{(1234)}
=\sum_{r=1}^{4}\Bigg\{&
 \frac{\mu_\ell^2\Pi_{abcd}}{\mathcal D_{01234}}
 -\frac{\Delta_{ab|cd}\Tcal_{ab|cd}}{\mathcal D_{024}}\\
&+\frac{\PT(1234)}{2}\bigg[
 \frac{s_{ab}s_{bc}(\bar R_r-R_r)
       +\tr_5(\bar\ell,p_a,p_b,p_c)(\bar R_r+R_r)}
      {\mathcal D_{0123}}\\
&\qquad
 +\frac{s_{abc}s_{cd}(\bar R_r-R_r)
       +\tr_5(\bar\ell,p_c,p_a+p_b,p_d)(\bar R_r+R_r)}
      {\mathcal D_{0234}}\\
&\qquad
 +\frac{s_{ab}s_{bcd}(\bar R_{r+1}-R_{r+1})
       -\tr_5(\bar\ell,p_a,p_b,p_c+p_d)(\bar R_{r+1}+R_{r+1})}
      {\mathcal D_{0124}}\\
&\qquad
 +\frac{(q^2-s_{bcd})(\bar R_{r+1}-R_{r+1})}{\mathcal D_{014}}
 +\frac{(q^2-s_{abc})(\bar R_r-R_r)}{\mathcal D_{034}}
 \bigg]\Bigg\}.
\end{aligned}
\label{eq:one-loop-integrand}
\end{equation}
In the $r$th term of the sum, $(a,b,c,d)=(r,r+1,r+2,r+3)$, with leg
labels and $R$-invariant indices taken modulo four.
The box numerators use the multilinear continuation of the oriented trace
\begin{equation}
\tr_5(p_1,p_2,p_3,p_4)
=[12]\langle23\rangle[34]\langle41\rangle
-\langle12\rangle[23]\langle34\rangle[41],
\end{equation}
in the convention $s_{ij}=\langle ij\rangle[ij]$.
The pentagon coefficient includes the Parke--Taylor factor,
\begin{equation}
\begin{aligned}
\Pi_{abcd}
&=-\frac{\epsilon_{ABCD}M_{abcd}^{AB}M_{abcd}^{CD}}
         {4!\,s_{bc}(s_{abc}s_{bcd}-q^2s_{bc})},\\
M_{abcd}^{AB}
&=\frac{d_{da}s_{bcd}\xi_{abc}^A\xi_{abc}^B
        +d_{ad}s_{abc}\xi_{bcd}^A\xi_{bcd}^B
        -2d_{ad}d_{da}\xi_{abc}^{[A}\xi_{bcd}^{B]}}
       {q^2[bc]},\\
d_{ij}&=\sum_{k=1}^4\langle jk\rangle[ik],
\qquad
\xi_{ijk}^A=[ij]\eta_k^A+[jk]\eta_i^A+[ki]\eta_j^A .
\end{aligned}
\label{eq:one-loop-pentagon}
\end{equation}
We take $\epsilon_{1234}=1$ and antisymmetrize with weight $1/2$.
The same formula for $\Pi_{abcd}$ applies to all four rotations.

For a cut propagator $\mathcal D_j$, the $D$-dimensional on-shell condition
reads
\begin{equation}
\ell=\bar\ell+\ell_\perp,\qquad \mu_\ell^2=-\ell_\perp^2,\qquad
(\ell-K_j)^2=0\quad\Longleftrightarrow\quad
(\bar\ell-K_j)^2=\mu_\ell^2 .
\label{eq:one-loop-cut-condition}
\end{equation}
In four dimensions this is the on-shell condition for a state of mass
squared $\mu_\ell^2$. We represent the transverse momentum by a constant
scalar background along $\ell_\perp$ in scalar space. The commutator with
the background gives the charged cut states this mass. Writing
$\phi=Y\cdot\Phi=\phi_{\rm bg}+\varphi$, with $Y$ the scalar polarization
and $\varphi$ the fluctuation, gives
\begin{equation*}
\tr(\phi^2)=\tr(\phi_{\rm bg}^2)+2\tr(\phi_{\rm bg}\varphi)
+\tr(\varphi^2).
\end{equation*}
The background carries zero momentum. The term linear in $\varphi$
therefore couples the operator to the tree through one off-shell scalar
carrying the full momentum $q$.
Let $k_a,k_b$ be the momenta of two bosonic cut legs, including their
transverse components, and $e_a,e_b$ their polarizations. The
color-stripped minimal tree form factor is
\begin{equation}
(Y\cdot e_a)(Y\cdot e_b)
-\frac{Y\cdot k_a}{q^2}\,Y\cdot V_3(k_a,e_a;k_b,e_b),
\qquad k_a+k_b=q .
\label{eq:one-loop-source-cut}
\end{equation}
For outgoing momenta $k_a,k_b,-q$, the color-stripped cubic current is
\begin{equation}
V_3^\mu(k_a,e_a;k_b,e_b)
=(k_a-k_b)^\mu(e_a\cdot e_b)
+e_b^\mu(k_b+q)\cdot e_a-e_a^\mu(k_a+q)\cdot e_b .
\label{eq:one-loop-cubic-current}
\end{equation}
Setting $e_a=k_a$ and using $k_a^2=k_b^2=k_b\cdot e_b=0$ gives
\begin{equation*}
V_3^\mu(k_a,k_a;k_b,e_b)=q^2 e_b^\mu-q^\mu(k_a\cdot e_b).
\end{equation*}
Since $Y\cdot q=0$, the two terms in Eq.~\eqref{eq:one-loop-source-cut}
cancel under this replacement. The same Ward identity holds for the
other cut leg.
We extract the $\mathrm{SU}(4)$-invariant Grassmann coefficients of the
cuts, retaining their dependence on $\mu_\ell^2$. The five-propagator cut
fixes $\Pi_{abcd}$ in Eq.~\eqref{eq:one-loop-integrand}, and the lower cuts
fix the box and triangle numerators. We check the integrand at two independent
finite-field points by averaging the directly evaluated cuts over the
transverse direction. We obtain the one-loop form factor through
$O(\eps^2)$ using the one-mass pentagon functions
\cite{AbreuItaMoriello:2020,Abreu:2024,Abreu:2023data}.
\par
\endgroup

\section{Generalized unitarity cuts}
\label{app:cut-topologies}

\begingroup
\tikzset{
  cut skeleton/.style={x=7.7mm,y=8.4mm,line width=0.5pt,line cap=round},
  operator leg/.style={double=white,double distance=1.15pt,line cap=butt}
}
\newcommand{\CutCell}[3][0.70]{%
  \makebox[.16\linewidth][c]{%
    \scalebox{#1}{%
      \begin{tikzpicture}[cut skeleton]
        \useasboundingbox (-2,-1.27) rectangle (2,1.27);
        \foreach \id/\vx/\vy in {#2}
          \coordinate (v\id) at (\vx,\vy);
        #3
        \foreach \id/\vx/\vy in {#2}
          \draw[fill=white] (v\id) circle[radius=3.2pt];
      \end{tikzpicture}%
    }%
  }\ignorespaces
}
\newcommand{\CutRow}[1]{\noindent\makebox[\linewidth][c]{#1}\par}
\makeatletter
\newcommand{\CutPanel}[3]{%
  \par\addvspace{8pt}\noindent
  \begin{minipage}{\linewidth}
    #3%
    \setlength{\abovecaptionskip}{2pt}%
    \def\@captype{figure}\caption{#2}\label{#1}%
  \end{minipage}\par
}
\makeatother

Each internal edge denotes a cut propagator, and each circle represents
a tree amplitude or a tree form factor. The double external line marks
the operator insertion. We classify the graphs up to permutations of the
four on-shell external legs, with the helicity sectors of the tree
factors suppressed.

The cuts in Figs.~\ref{fig:cuts-eight}--\ref{fig:cuts-seven-two-external}
and \ref{fig:cuts-six} fix the remaining symbol coefficients.
Figure~\ref{fig:cuts-four} shows a four-propagator cut that provides an
independent check of the reconstructed symbol.
The seven-propagator cuts in Fig.~\ref{fig:cuts-seven-form-factor}
also enter the matching through the finite term.

\CutPanel{fig:cuts-eight}{Eight-propagator cuts.}{%
\CutRow{%
\CutCell{0/-1.37/0.68, 1/0/0.68, 2/-1.37/-0.68, 3/0/-0.68, 4/1.37/-0.68, 5/1.37/0.68, 6/0/0}{
  \draw (v1) -- (v0) -- (v2) -- (v3);
  \draw (v1) -- (v6) -- (v3);
  \draw (v1) -- (v5) -- (v4) -- (v3);
  \draw[operator leg] (v0) -- ++(-0.518,0.323);
  \draw (v2) -- ++(-0.424,-0.265);
  \draw (v4) -- ++(0.424,-0.265);
  \draw (v5) -- ++(0.424,0.265);
  \draw (v6) -- ++(-0.43,0);
}
\CutCell{0/-1.37/0.68, 1/0/0.68, 2/-1.37/-0.68, 3/0/-0.68, 4/1.37/-0.68, 5/1.37/0.68, 6/0/0}{
  \draw (v1) -- (v0) -- (v2) -- (v3);
  \draw (v1) -- (v6) -- (v3);
  \draw (v1) -- (v5) -- (v4) -- (v3);
  \draw (v0) -- ++(-0.424,0.265);
  \draw (v2) -- ++(-0.424,-0.265);
  \draw (v4) -- ++(0.424,-0.265);
  \draw (v5) -- ++(0.424,0.265);
  \draw[operator leg] (v6) -- ++(-0.61,0);
}
\CutCell{0/-1.21/0.68, 1/0/0.68, 2/-1.46/0, 3/-1.21/-0.68, 4/0/-0.68, 5/0/0, 6/1.18/0}{
  \draw (v1) -- (v0) -- (v2) -- (v3) -- (v4);
  \draw (v1) -- (v5) -- (v4);
  \draw (v1) -- (v6) -- (v4);
  \draw[operator leg] (v0) -- ++(-0.518,0.323);
  \draw (v2) -- ++(-0.5,0);
  \draw (v3) -- ++(-0.424,-0.265);
  \draw (v5) -- ++(-0.43,0);
  \draw (v6) -- ++(0.5,0);
}
\CutCell{0/-1.21/0.68, 1/0/0.68, 2/-1.46/0, 3/-1.21/-0.68, 4/0/-0.68, 5/0/0, 6/1.18/0}{
  \draw (v1) -- (v0) -- (v2) -- (v3) -- (v4);
  \draw (v1) -- (v5) -- (v4);
  \draw (v1) -- (v6) -- (v4);
  \draw (v0) -- ++(-0.424,0.265);
  \draw[operator leg] (v2) -- ++(-0.61,0);
  \draw (v3) -- ++(-0.424,-0.265);
  \draw (v5) -- ++(-0.43,0);
  \draw (v6) -- ++(0.5,0);
}
}
\CutRow{%
\CutCell{0/1.21/0.68, 1/0/0.68, 2/1.46/0, 3/1.21/-0.68, 4/0/-0.68, 5/0/0, 6/-1.18/0}{
  \draw (v1) -- (v6) -- (v4);
  \draw (v1) -- (v5) -- (v4);
  \draw (v1) -- (v0) -- (v2) -- (v3) -- (v4);
  \draw (v0) -- ++(0.424,0.265);
  \draw (v2) -- ++(0.5,0);
  \draw (v3) -- ++(0.424,-0.265);
  \draw (v5) -- ++(-0.43,0);
  \draw[operator leg] (v6) -- ++(-0.61,0);
}
\CutCell{0/-1.21/0.68, 1/0/0.68, 2/-1.46/0, 3/-1.21/-0.68, 4/0/-0.68, 5/1.37/0.68, 6/1.37/-0.68}{
  \draw (v1) -- (v0) -- (v2) -- (v3) -- (v4);
  \draw (v1) -- (v4);
  \draw (v1) -- (v5) -- (v6) -- (v4);
  \draw[operator leg] (v0) -- ++(-0.518,0.323);
  \draw (v2) -- ++(-0.5,0);
  \draw (v3) -- ++(-0.424,-0.265);
  \draw (v5) -- ++(0.424,0.265);
  \draw (v6) -- ++(0.424,-0.265);
}
\CutCell{0/-1.21/0.68, 1/0/0.68, 2/-1.46/0, 3/-1.21/-0.68, 4/0/-0.68, 5/1.37/0.68, 6/1.37/-0.68}{
  \draw (v1) -- (v0) -- (v2) -- (v3) -- (v4);
  \draw (v1) -- (v4);
  \draw (v1) -- (v5) -- (v6) -- (v4);
  \draw (v0) -- ++(-0.424,0.265);
  \draw[operator leg] (v2) -- ++(-0.61,0);
  \draw (v3) -- ++(-0.424,-0.265);
  \draw (v5) -- ++(0.424,0.265);
  \draw (v6) -- ++(0.424,-0.265);
}
\CutCell{0/1.21/0.68, 1/0/0.68, 2/1.46/0, 3/1.21/-0.68, 4/0/-0.68, 5/-1.37/0.68, 6/-1.37/-0.68}{
  \draw (v1) -- (v5) -- (v6) -- (v4);
  \draw (v1) -- (v4);
  \draw (v1) -- (v0) -- (v2) -- (v3) -- (v4);
  \draw (v0) -- ++(0.424,0.265);
  \draw (v2) -- ++(0.5,0);
  \draw (v3) -- ++(0.424,-0.265);
  \draw[operator leg] (v5) -- ++(-0.518,0.323);
  \draw (v6) -- ++(-0.424,-0.265);
}
}
}
\CutPanel{fig:cuts-seven-one-external}{Seven-propagator cuts with one external leg on a four-point tree amplitude or with all three legs of a three-point tree form factor cut.}{%
\CutRow{%
\CutCell{0/0/0, 1/0/-0.68, 2/-1.37/-0.68, 3/-1.37/0.68, 4/1.18/0, 5/0/0.68}{
  \draw (v5) -- (v3) -- (v2) -- (v1);
  \draw (v5) -- (v0) -- (v1);
  \draw (v5) -- (v4) -- (v1);
  \draw (v0) -- ++(-0.43,0);
  \draw (v2) -- ++(-0.424,-0.265);
  \draw (v3) -- ++(-0.424,0.265);
  \draw (v4) -- ++(0.5,0);
  \draw[operator leg] (v5) -- ++(0,0.61);
}
\CutCell{0/-1.18/0, 1/0/0.68, 2/1.37/-0.68, 3/1.37/0.68, 4/0/0, 5/0/-0.68}{
  \draw (v1) -- (v0) -- (v5);
  \draw (v1) -- (v4) -- (v5);
  \draw (v1) -- (v3) -- (v2) -- (v5);
  \draw[operator leg] (v0) -- ++(-0.61,0);
  \draw (v2) -- ++(0.424,-0.265);
  \draw (v3) -- ++(0.424,0.265);
  \draw (v4) -- ++(-0.43,0);
  \draw (v5) -- ++(0,-0.5);
}
\CutCell{0/-1.37/0.68, 1/0/0.68, 2/-1.37/-0.68, 3/0/0, 4/1.18/0, 5/0/-0.68}{
  \draw (v1) -- (v0) -- (v2) -- (v5);
  \draw (v1) -- (v3) -- (v5);
  \draw (v1) -- (v4) -- (v5);
  \draw[operator leg] (v0) -- ++(-0.518,0.323);
  \draw (v2) -- ++(-0.424,-0.265);
  \draw (v3) -- ++(-0.43,0);
  \draw (v4) -- ++(0.5,0);
  \draw (v5) -- ++(0,-0.5);
}
\CutCell{0/-1.37/0.68, 1/-1.37/-0.68, 2/0/-0.68, 3/0/0, 4/1.18/0, 5/0/0.68}{
  \draw (v5) -- (v0) -- (v1) -- (v2);
  \draw (v5) -- (v3) -- (v2);
  \draw (v5) -- (v4) -- (v2);
  \draw[operator leg] (v0) -- ++(-0.518,0.323);
  \draw (v1) -- ++(-0.424,-0.265);
  \draw (v3) -- ++(-0.43,0);
  \draw (v4) -- ++(0.5,0);
  \draw (v5) -- ++(0,0.5);
}
\CutCell{0/-1.37/0.68, 1/-1.37/-0.68, 2/0/-0.68, 3/1.37/-0.68, 4/1.37/0.68, 5/0/0.68}{
  \draw (v5) -- (v0) -- (v1) -- (v2);
  \draw (v5) -- (v2);
  \draw (v5) -- (v4) -- (v3) -- (v2);
  \draw[operator leg] (v0) -- ++(-0.518,0.323);
  \draw (v1) -- ++(-0.424,-0.265);
  \draw (v3) -- ++(0.424,-0.265);
  \draw (v4) -- ++(0.424,0.265);
  \draw (v5) -- ++(0,0.5);
}
\CutCell{0/-1.37/0.68, 1/0/0.68, 2/-1.37/-0.68, 3/1.37/-0.68, 4/1.37/0.68, 5/0/-0.68}{
  \draw (v1) -- (v0) -- (v2) -- (v5);
  \draw (v1) -- (v5);
  \draw (v1) -- (v4) -- (v3) -- (v5);
  \draw[operator leg] (v0) -- ++(-0.518,0.323);
  \draw (v2) -- ++(-0.424,-0.265);
  \draw (v3) -- ++(0.424,-0.265);
  \draw (v4) -- ++(0.424,0.265);
  \draw (v5) -- ++(0,-0.5);
}
}
}

\CutPanel{fig:cuts-seven-two-external}{Seven-propagator cuts with two external legs on a four-point tree amplitude.}{%
\CutRow{%
\CutCell{0/-1.37/0.68, 1/0/0.68, 2/-1.37/-0.68, 3/0/-0.68, 4/0/0, 5/1.18/0}{
  \draw (v1) -- (v0) -- (v2) -- (v3);
  \draw (v1) -- (v4) -- (v3);
  \draw (v1) -- (v5) -- (v3);
  \draw[operator leg] (v0) -- ++(-0.518,0.323);
  \draw (v2) -- ++(-0.424,-0.265);
  \draw (v4) -- ++(-0.43,0);
  \draw (v5) -- ++(0.467,-0.179);
  \draw (v5) -- ++(0.467,0.179);
}
\CutCell{0/0/0.68, 1/0/-0.68, 2/1.37/-0.68, 3/1.37/0.68, 4/-1.18/0, 5/0/0}{
  \draw (v0) -- (v4) -- (v1);
  \draw (v0) -- (v5) -- (v1);
  \draw (v0) -- (v3) -- (v2) -- (v1);
  \draw (v2) -- ++(0.424,-0.265);
  \draw (v3) -- ++(0.424,0.265);
  \draw[operator leg] (v4) -- ++(-0.61,0);
  \draw (v5) -- ++(-0.401,0.154);
  \draw (v5) -- ++(-0.401,-0.154);
}
\CutCell{0/-1.37/0.68, 1/0/0.68, 2/0/-0.68, 3/0/0, 4/1.18/0, 5/-1.37/-0.68}{
  \draw (v1) -- (v0) -- (v5) -- (v2);
  \draw (v1) -- (v3) -- (v2);
  \draw (v1) -- (v4) -- (v2);
  \draw[operator leg] (v0) -- ++(-0.518,0.323);
  \draw (v3) -- ++(-0.43,0);
  \draw (v4) -- ++(0.5,0);
  \draw (v5) -- ++(-0.491,-0.095);
  \draw (v5) -- ++(-0.301,-0.399);
}
\CutCell{0/0/0.68, 1/1.37/-0.68, 2/0/-0.68, 3/0/0, 4/-1.18/0, 5/1.37/0.68}{
  \draw (v0) -- (v4) -- (v2);
  \draw (v0) -- (v3) -- (v2);
  \draw (v0) -- (v5) -- (v1) -- (v2);
  \draw (v1) -- ++(0.424,-0.265);
  \draw (v3) -- ++(-0.43,0);
  \draw[operator leg] (v4) -- ++(-0.61,0);
  \draw (v5) -- ++(0.491,0.095);
  \draw (v5) -- ++(0.301,0.399);
}
\CutCell{0/-1.37/0.68, 1/0/0.68, 2/0/-0.68, 3/1.37/0.68, 4/1.37/-0.68, 5/-1.37/-0.68}{
  \draw (v1) -- (v0) -- (v5) -- (v2);
  \draw (v1) -- (v2);
  \draw (v1) -- (v3) -- (v4) -- (v2);
  \draw[operator leg] (v0) -- ++(-0.518,0.323);
  \draw (v3) -- ++(0.424,0.265);
  \draw (v4) -- ++(0.424,-0.265);
  \draw (v5) -- ++(-0.491,-0.095);
  \draw (v5) -- ++(-0.301,-0.399);
}
}
\CutRow{%
\CutCell{0/-1.21/0.68, 1/0/0.68, 2/-1.46/0, 3/-1.21/-0.68, 4/0/-0.68, 5/1.18/0}{
  \draw (v1) -- (v0) -- (v2) -- (v3) -- (v4);
  \draw (v1) -- (v4);
  \draw (v1) -- (v5) -- (v4);
  \draw[operator leg] (v0) -- ++(-0.518,0.323);
  \draw (v2) -- ++(-0.5,0);
  \draw (v3) -- ++(-0.424,-0.265);
  \draw (v5) -- ++(0.467,-0.179);
  \draw (v5) -- ++(0.467,0.179);
}
\CutCell{0/-1.21/0.68, 1/0/0.68, 2/-1.46/0, 3/-1.21/-0.68, 4/0/-0.68, 5/1.18/0}{
  \draw (v1) -- (v0) -- (v2) -- (v3) -- (v4);
  \draw (v1) -- (v4);
  \draw (v1) -- (v5) -- (v4);
  \draw (v0) -- ++(-0.424,0.265);
  \draw[operator leg] (v2) -- ++(-0.61,0);
  \draw (v3) -- ++(-0.424,-0.265);
  \draw (v5) -- ++(0.467,-0.179);
  \draw (v5) -- ++(0.467,0.179);
}
\CutCell{0/0/0.68, 1/1.37/-0.68, 2/0/-0.68, 3/-1.37/0.68, 4/-1.37/-0.68, 5/1.37/0.68}{
  \draw (v0) -- (v3) -- (v4) -- (v2);
  \draw (v0) -- (v2);
  \draw (v0) -- (v5) -- (v1) -- (v2);
  \draw (v1) -- ++(0.424,-0.265);
  \draw[operator leg] (v3) -- ++(-0.518,0.323);
  \draw (v4) -- ++(-0.424,-0.265);
  \draw (v5) -- ++(0.491,0.095);
  \draw (v5) -- ++(0.301,0.399);
}
\CutCell{0/1.37/0.68, 1/0/0.68, 2/0/-0.68, 3/-1.37/0.68, 4/-1.37/-0.68, 5/1.37/-0.68}{
  \draw (v1) -- (v3) -- (v4) -- (v2);
  \draw (v1) -- (v2);
  \draw (v1) -- (v0) -- (v5) -- (v2);
  \draw (v0) -- ++(0.424,0.265);
  \draw[operator leg] (v3) -- ++(-0.518,0.323);
  \draw (v4) -- ++(-0.424,-0.265);
  \draw (v5) -- ++(0.301,-0.399);
  \draw (v5) -- ++(0.491,-0.095);
}
}
}
\CutPanel{fig:cuts-seven-form-factor}{Seven-propagator cuts with one external leg on a three-point tree form factor.}{%
\CutRow{%
\CutCell{0/0/0.68, 1/0/-0.68, 2/1.37/-0.68, 3/1.37/0.68, 4/0/0, 5/-1.18/0}{
  \draw (v0) -- (v5) -- (v1);
  \draw (v0) -- (v4) -- (v1);
  \draw (v0) -- (v3) -- (v2) -- (v1);
  \draw (v2) -- ++(0.424,-0.265);
  \draw (v3) -- ++(0.424,0.265);
  \draw (v4) -- ++(-0.43,0);
  \draw[operator leg] (v5) -- ++(-0.569,0.219);
  \draw (v5) -- ++(-0.467,-0.179);
}
\CutCell{0/0/0.68, 1/-1.37/-0.68, 2/0/-0.68, 3/0/0, 4/1.18/0, 5/-1.37/0.68}{
  \draw (v0) -- (v5) -- (v1) -- (v2);
  \draw (v0) -- (v3) -- (v2);
  \draw (v0) -- (v4) -- (v2);
  \draw (v1) -- ++(-0.424,-0.265);
  \draw (v3) -- ++(-0.43,0);
  \draw (v4) -- ++(0.5,0);
  \draw[operator leg] (v5) -- ++(-0.367,0.487);
  \draw (v5) -- ++(-0.491,0.095);
}
\CutCell{0/0/0.68, 1/-1.37/-0.68, 2/0/-0.68, 3/1.37/0.68, 4/1.37/-0.68, 5/-1.37/0.68}{
  \draw (v0) -- (v5) -- (v1) -- (v2);
  \draw (v0) -- (v2);
  \draw (v0) -- (v3) -- (v4) -- (v2);
  \draw (v1) -- ++(-0.424,-0.265);
  \draw (v3) -- ++(0.424,0.265);
  \draw (v4) -- ++(0.424,-0.265);
  \draw[operator leg] (v5) -- ++(-0.367,0.487);
  \draw (v5) -- ++(-0.491,0.095);
}
}
}

\CutPanel{fig:cuts-six}{Six-propagator cuts.}{%
\CutRow{%
\CutCell{0/1.18/0, 1/-1.37/0.68, 2/-1.37/-0.68, 3/0/0.68, 4/0/-0.68}{
  \draw (v3) -- (v1) -- (v2) -- (v4);
  \draw (v3) -- (v4);
  \draw (v3) -- (v0) -- (v4);
  \draw (v0) -- ++(0.5,0);
  \draw (v1) -- ++(-0.424,0.265);
  \draw (v2) -- ++(-0.424,-0.265);
  \draw[operator leg] (v3) -- ++(0,0.61);
  \draw (v4) -- ++(0,-0.5);
}
\CutCell{0/-1.37/0.68, 1/-1.37/-0.68, 2/1.18/0, 3/0/-0.68, 4/0/0.68}{
  \draw (v4) -- (v0) -- (v1) -- (v3);
  \draw (v4) -- (v3);
  \draw (v4) -- (v2) -- (v3);
  \draw[operator leg] (v0) -- ++(-0.518,0.323);
  \draw (v1) -- ++(-0.424,-0.265);
  \draw (v2) -- ++(0.5,0);
  \draw (v3) -- ++(0,-0.5);
  \draw (v4) -- ++(0,0.5);
}
\CutCell{0/-1.21/0.68, 1/-1.46/0, 2/-1.21/-0.68, 3/0/0.68, 4/0/-0.68}{
  \draw (v3) -- (v0) -- (v1) -- (v2) -- (v4);
  \draw (v3) -- (v4);
  \draw (v3) .. controls (1.27,0.68) and (1.27,-0.68) .. (v4);
  \draw[operator leg] (v0) -- ++(-0.518,0.323);
  \draw (v1) -- ++(-0.5,0);
  \draw (v2) -- ++(-0.424,-0.265);
  \draw (v3) -- ++(0,0.5);
  \draw (v4) -- ++(0,-0.5);
}
\CutCell{0/-1.18/0, 1/1.37/0.68, 2/1.37/-0.68, 3/0/0.68, 4/0/-0.68}{
  \draw (v3) -- (v0) -- (v4);
  \draw (v3) -- (v4);
  \draw (v3) -- (v1) -- (v2) -- (v4);
  \draw[operator leg] (v0) -- ++(-0.61,0);
  \draw (v1) -- ++(0.424,0.265);
  \draw (v2) -- ++(0.424,-0.265);
  \draw (v3) -- ++(0,0.5);
  \draw (v4) -- ++(0,-0.5);
}
\CutCell{0/-1.21/0.68, 1/-1.46/0, 2/-1.21/-0.68, 3/0/0.68, 4/0/-0.68}{
  \draw (v3) -- (v0) -- (v1) -- (v2) -- (v4);
  \draw (v3) -- (v4);
  \draw (v3) .. controls (1.27,0.68) and (1.27,-0.68) .. (v4);
  \draw (v0) -- ++(-0.424,0.265);
  \draw[operator leg] (v1) -- ++(-0.61,0);
  \draw (v2) -- ++(-0.424,-0.265);
  \draw (v3) -- ++(0,0.5);
  \draw (v4) -- ++(0,-0.5);
}
\CutCell{0/-1.21/0.68, 1/-1.46/0, 2/-1.21/-0.68, 3/0/0.68, 4/0/-0.68}{
  \draw (v3) -- (v0) -- (v1) -- (v2) -- (v4);
  \draw (v3) -- (v4);
  \draw (v3) .. controls (1.27,0.68) and (1.27,-0.68) .. (v4);
  \draw (v0) -- ++(-0.424,0.265);
  \draw (v1) -- ++(-0.5,0);
  \draw (v2) -- ++(-0.424,-0.265);
  \draw[operator leg] (v3) -- ++(0,0.61);
  \draw (v4) -- ++(0,-0.5);
}
}
}
\CutPanel{fig:cuts-four}{Four-propagator cut in the $s_{12}$ and $q^2$ channels.}{%
\CutRow{%
\CutCell[1.0]{0/-1.22/0, 1/0/0, 2/1.22/0}{
  \draw (v0) to[bend left=38] (v1) to[bend left=38] (v2);
  \draw (v0) to[bend right=38] (v1) to[bend right=38] (v2);
  \draw (v0) -- ++(-.44,.37) (v0) -- ++(-.44,-.37);
  \draw (v1) -- ++(0,.71) (v1) -- ++(0,-.71);
  \draw[operator leg] (v2) -- ++(.61,0);
}
}
}
\par
\endgroup

\end{document}